\documentclass{aa} 
\usepackage{txfonts}  
\usepackage{graphicx} 
\usepackage[figuresright]{rotating}
\usepackage{url}
\usepackage{longtable,lscape}
\usepackage{natbib} 
\usepackage{epsfig}
\usepackage[ps2pdf]{thumbpdf}
\usepackage[bookmarks]{hyperref}
\usepackage{acronym}
\usepackage[normalem]{ulem}
\usepackage{float}
\newcommand{\snm} {V351~Ori}
\newcommand{\ha}{$H\alpha$}
\newcommand{\hb}{$H\beta$}
\newcommand{\ndo}{$NaD1$} 
\newcommand{\ndt}{$NaD2$}
\newcommand{\near}{$\sim$}
\newcommand{\ang}{\AA}
\newcommand{\kms}{km~s$^{-1}$}
\newcommand{\ms}{m~s$^{-2}$}

\newcommand{\mult}{$\times$}  
\newcommand{\msun}{$M_\odot$}
\newcommand{\rsun}{$R_\odot$}

\newcommand{\ewha}{EW$_{\lambda}$[\ha\ ]}
\newcommand\micron{\mbox{$\mu$m}}%
\begin{document}
\acrodef{SED}{spectral energy distribution}
\acrodef{HAeBe}{Herbig Ae/Be}
\acrodef{NIR}{near infrared}
\acrodef{FIR}{far infrared}
\acrodef{YSO}{young stellar objects}
\acrodef{PMS}{pre-main sequence}
\acrodef{IRAS}{Infrared Astronomical Satellite}
\acrodef{TACs}{transient absorption components}
\acrodef{CTTS}{classical T Tauri stars}
%
\title{Variable circumstellar activity of V351 Orionis}
\author{Rumpa Choudhury, H. C. Bhatt,  Gajendra Pandey}
\institute{Indian Institute of Astrophysics, Bangalore 560034, India}
\offprints{R. Choudhury, \email rumpa@iiap.res.in}  
\date{Received /Accepted}  
\titlerunning{Variable circumstellar activity of \snm}  
\authorrunning{R. Choudhury et. al.}  
\abstract 
{Emission  and  absorption  line  profiles  which are  formed  in  the
  interaction  of \ac{PMS} stars  and their  circumstellar environment
  are found to be variable at various timescales.}
{We investigate  the patterns and timescales of  temporal line profile
  variability  in   order  to  explore   the  dynamical  circumstellar
  environment of the \ac{PMS} Herbig~Ae star \snm.}
{We  obtained 45  high-resolution (R\near28,000)  spectra of  \snm\ at
  timescales of  hours, days,  and months. We  analysed the  \ha\ line
  profiles and also examined the \hb, \ndo\ and \ndt\ line profiles to
  explore the nature of the spectroscopic variability.}
{The   \ha\  line   profiles   showed  strong   variations  over   all
  timescales. The shape  of the profiles changed over  timescales of a
  day.  Single  as well as  simultaneous event(s) of  blue-shifted and
  red-shifted \ac{TACs}, i.e.  signatures  of outflow and infall, were
  also observed  in the  \ha\ line profiles.   The shortest  period of
  variation  in  the  \ac{TACs}  was  $\leq$ 1  hour.   All  transient
  absorption events were  found to decelerate with a rate  of a few to
  fractions of m  s$^{-2}$. The depth and width  of the \ac{TACs} were
  also  changing with  time.   The presence  of elongated  red-shifted
  components  at   some  epochs   supports  the  episodic   nature  of
  accretion.}
{Variable  emission and  absorption components  detected in  \ha\ line
  profiles show  the dynamic nature  of interaction between  \snm\ and
  its circumstellar  environment.  The \ha\  non-photospheric profiles
  of  the star  most probably  originate in  the disk  wind.  Episodic
  accretion of gaseous  material at a slow rate  and outflow of clumpy
  gaseous material are still occurring in \snm\ at an age of $\sim$6.5
  Myr.  Dynamic  magnetospheric accretion and disk wind  emerge as the
  most satisfactory  model for interpreting the  observed line profile
  variations of \snm.}
%
\keywords{stars:pre   main  sequence,   (stars):circumstellar  matter,
  stars:individual:  \snm, techniques:spectroscopic,  line:profiles}
\maketitle
\section{Introduction} 
\label{section:intro} 
\ac{HAeBe} stars  are revealed  pre-main sequence (PMS)  emission line
stars  with  masses of  \near1.5  to 10  \msun\  and  typical ages  of
\near$10^6$  years  with  spectral  types  earlier  than  \textit{F0}.
Balmer emission  lines of hydrogen  and infrared (IR)  excess emission
produced  by thermal  emission of  circumstellar dust  grains  are the
prominent  characteristics   of  \ac{HAeBe}  stars.    Both  of  these
signatures originate in the  circumstellar material around these stars
and the interaction  between them.  It has been  well established that
circumstellar  material,  i.e.  dust  and  gas  around the  \ac{HAeBe}
stars,  are  distributed   in  disks  (e.g.   \cite{maheswar2002}  and
references therein).  Significant variations in shape and intensity of
permitted  and forbidden  emission line  profiles, originating  in the
accretion  or outflow events  of the  interaction between  the central
star  and its surrounding  material, are  observed in  many \ac{HAeBe}
stars    \citep{hamann1992,   hamann1994,    reipurth1996}.    Optical
spectroscopic studies  of emission line  profiles of a  few \ac{HAeBe}
stars  are  reported  in  the literature  in  e.g.   \cite{grinin1994,
dewinter1999,     grinin2001,    natta2000,     mora2002,    mora2004,
guimaraes2006}.  \cite{natta2000}  reported the evidence  for episodic
rather   than   continuous  accretion   in   Herbig~Ae  star   UX~Ori.
\cite{mora2002,  mora2004}  reported  transient absorption  components
(TACs) in  several Balmer and metallic  lines of a  few Herbig~Ae type
stars, which are supposed to be created by moving gaseous blobs around
these  stars.   These studies  discussed  the dynamical  circumstellar
environment around \ac{HAeBe} stars which show strong \ha\ emission.
 
\cite{manoj06} showed  that the  emission line activity  of \ac{HAeBe}
stars substantially  decreases by  \near3~Myr, which agrees  well with
the  inner disk  survival  timescale i.e  \near3~Myr  as suggested  by
\cite{hernandez2005, hernandez2009} for  the early spectral type stars
such as B,  A, and F types.  The dissipation  of optically thick disks
beyond \near3~Myr  to the  onset of young  main sequence stars  with a
debris disk can be caused  by various disk dispersal processes such as
photoevaporation, planet  formation, etc.  \ac{HAeBe}  stars with weak
emission  lines  and moderate  IR  excess  can  be considered  as  the
precursor  of the  young main  sequence stars  with IR  excess  and no
gaseous    emission   lines,    e.g.     \textit{$\beta$   Pic}    and
\textit{Vega}. Understanding the nature of accretion and circumstellar
material of  comparatively less active  \ac{HAeBe} stars of  age $\ge$
3~Myr are  important because these processes eventually  set the stage
for planet formation. However,  detailed studies of weak emission line
\ac{HAeBe} stars that cover hours,  days, and months are not available
in the literature.

V351  Orionis (\snm\,, HIP~27059,  HD~38238) is  a weak  \ha\ emission
line  Herbig~Ae   star  that  showed   photometric  and  spectroscopic
variabilities over  various timescales in  e.g.  \cite{ vanancker1996,
  balona2002} and references therein.   The classification of \snm\ as
a  Herbig  Ae  star  has   been  confirmed  by  several  authors,  e.g
\cite{vanancker1996,   vieira2003,  hernandez2005}.   \cite{grady1996}
analysed the  ultraviolet (UV) spectra of  \snm\ and did  not find any
absorption component,  i.e.  signatures of stellar  wind or accretion,
in  the  spectra.    \cite{vanancker1996}  presented  photometric  and
spectroscopic observations of \snm\ and found that it transformed into
an almost non-variable star  from a strong photometric variable within
a short period  of \near14 years.  \citeauthor{vanancker1996} reported
that the \ha\ profile of \snm\ was of an inverse P Cygni type and also
showed a high infrared excess  emission.  Inverse P Cygni profiles are
signatures of the accretion of circumstellar material onto the central
object and are characterised by blue-shifted emission with red-shifted
absorptions  or  systematically   enhanced  red  shifted  absorptions.
\snm\ shows $\delta$  Scuti type pulsations \citep{marconi2000}, which
are  another   characteristic  of   \ac{PMS}  stars  of   mass  $\geq$
1.5\msun\  as  they  cross  the pulsation  instability  towards  their
contraction   to  the   main  sequence.    \cite{balona2002}  obtained
simultaneous  multi-colour photometric and  spectroscopic observations
of  \snm\  to  investigate  the  nature  of  the  pulsation  of  \snm.
\citeauthor{balona2002} also reported a  variable inverse P Cygni type
\ha\ profile  with a variation on  a timescale of a  day with complete
absence  of emission  at some  epochs.  \cite{vieira2003}  observed an
inverse  P Cygni type  \ha\ emission  line profile  in \snm\  and also
suggested that  \snm\ might be  an evolved Herbig~Ae star,  because it
shows weak  emission lines.  \cite{hernandez2005} also  reported a low
equivalent  width  of  \ha\  i.e  \ewha\ =--0.9  \AA\  and  associated
\snm\  with Ori~OB1bc  region.   These signatures  hint  at an  active
interaction of the  weak \ha\ emission line young  star \snm\ with its
circumstellar environment.

In this work we  present high-resolution spectroscopic observations of
\snm\ along  with optical \textit{BVRI} observations.   We discuss the
dynamic circumstellar gaseous environment of  the star in the light of
available  models  of star-disk  interaction.   Observations and  data
reduction    are    discussed    in    Sect.~\ref{section:obs}.     In
Sect.~\ref{section:results} we present the  details of the spectra and
the analysis of  the \ac{TACs} detected in the  spectra.  A discussion
of the kinematics  of the observed \ac{TACs}, an  estimate of the disk
mass  and  a  qualitative  interpretation  of the  \ha\  line  profile
variation are  presented in Sect.~\ref{section:diss}.   We present our conclusion in
Sect.~\ref{section:con}.
%
\section{Observations and data reduction}
\label{section:obs}
Fortyfive  high-resolution  spectra  of  \snm\  were  obtained  during
October~2008  to April~2009 with  a fibre-fed  cross-dispersed echelle
spectrometer \citep{nkrao2005}  at the 2.3 m Vainu  Bappu Telescope of
Vainu    Bappu    Observatory    (VBO),    Kavalur,    India\footnote{
\url{http://www.iiap.res.in/vbo_vbt}}.  The log of the observations is
given   in  Table~\ref{tab_obslog}.   The   resolving  power   of  the
spectrometer  set-up  used  for  observation is  \near28,000  and  the
attached  2K\mult4K CCD  gives  a dispersion  of \near0.025~\ang\  per
pixel.   The  typical  integration   time  of  each  spectrum  was  45
minutes.  Bias subtraction,  flat field  correction and  scatter light
removal were carried  out for all the spectra  with the standard tasks
available   in    the   Image   Reduction    and   Analysis   Facility
(IRAF)\footnote{The  IRAF  software  is  distributed by  the  National
Optical  Astronomy  Observatory under  contact  with National  Science
Foundation.   \url{http://iraf.noao.edu/}}.  The  Th-Ar lamp  was used
for wavelength calibration.  The final  spectra that were used for our
analysis  cover  the wavelength  region  from  $\lambda$4250 \ang\  to
$\lambda$7900 \ang.   The CCD used  for observation did not  cover the
full  echelogram, and  as a  consequence there  are some  gaps  in the
wavelength coverage  of the  observed spectra.  However,  we optimised
the CCD position  accordingly to cover the wavelength  regions of \ha,
\hb, \ndo,  \ndt\ and  several other photospheric  lines.  Atmospheric
lines  were also  used  to check  the  wavelength calibration.   These
wavelength  calibrations  of the  final  spectra  are  accurate up  to
$\pm$0.01~\ang.

\snm\  was observed on  11~November~2009 with  the Bessell  broad band
filters { \em B}~(5\mult5~s), {\em V}~(4\mult2~s), {\em R}~(4\mult1~s)
and {\em  I}~( 5\mult1~s) of  the Himalayan Faint  Object Spectrograph
Camera (HFOSC) mounted on the  2m Himalayan Chandra Telescope (HCT) of
the  Indian  Astronomical  Observatory (IAO),  Hanle,  India\footnote{
\url{http://www.iiap.res.in/centers/iao}}.  Data were reduced with the
standard tasks in IRAF.

\begin{table}
\caption{Observation log of the echelle spectroscopy of \snm}
\label{tab_obslog}
\begin{tabular*}{\columnwidth}{cccc}
\hline
Date & JD -- 2454000  & t$_{exp}$(s) &\ha \, profile type$^{a}$ \\
& & &\\
\hline
\hline
2008-10-27 & 767.28 -- 767.41 & 2700 x 4 & IV-R,inverse P~Cygni\\
2008-10-30 & 770.38 & 2700 x 1 & II-R, double peak \\
2008-12-11 & 812.32 & 2700 x 1 & IV-R \\
2008-12-28 & 829.10 -- 829.32 & 2700 x 6 & IV-R\\
2008-12-29 & 830.10 -- 830.37 & 2700 x 8 & IV-R\\
2009-01-17 & 849.11 -- 849.22 & 2700 x 4 & III-R, double peak \\
2009-01-18 & 850.10 -- 850.32 & 2700 x 4 & III-R\\
2009-01-19 & 851.09 -- 851.35 & 2700 x 5 & III-R\\
2009-01-20 & 852.07 -- 852.18 & 2700 x 2 & III-R\\
2009-03-02 & 893.11 -- 893.15 & 2700 x 2 & III-R\\
2009-03-03 & 894.20 & 2700 x 1 & II-R\\
2009-03-29 & 920.16 & 2700 x 1 & IV-R\\
2009-03-30 & 921.07 -- 921.11 & 2700 x 2 & II-R\\
2009-03-31 & 922.08 & 2700 x 1 & III-R\\
2009-04-26 & 948.07 & 2700 x 1 & III-R\\
2009-04-27 & 949.07 & 2700 x 1 & II-R\\
2009-04-28 & 950.07 & 2700 x 1 & II-R\\
\hline
\end{tabular*}
{\noindent
\begin{tiny} a. Based on the classification scheme adopted from \cite{reipurth1996}. Type II-R profile represents a double peak profile where the red-sided peak exceeds  the strength of the blue-sided peak by half. Type III-R also represents double peak profile with the red-sided peak  less than half the strength of the blue-sided peak. Type IV-R represents the inverse
P Cygni profile. \end{tiny}\\}
\end{table}
%
\section{Results and analysis}
\label{section:results}
\subsection{Description of echelle spectra}
\ac{HAeBe} stars show both short  and long term variabilities in their
spectra.  To  explore the  short to long  term variations of  the line
profiles we obtained high-resolution spectra on timescales as short as
an hour to the longest coverage of \near7 months. All 45 spectra cover
the  wavelength regions  of \ha,  \hb, \ndo,  \ndt\ and  several other
photospheric lines.  Typical signal-  to-noise ratios (S/N) are $<$~10
to 12 at \hb, 15 to 30 at  \ndo\ and \ndt\ and $\geq$~30 to 60 at \ha.
We  note that [O~I]  $\lambda$6300~\AA\ emission  line was  present in
several spectra.   \cite{balona2002} also reported the  detection of a
narrow  emission line of  [O~I] $\lambda$6300~\ang\  at a  velocity of
+7.5 \kms.   However, the line centre of  [O~I] $\lambda$6300~\ang\ in
our observed  spectra remained  constant at laboratory  wavelength, as
did  the  other telluric  lines  throughout  our  observing runs.   We
therefore conclude that the [O~I] $\lambda$6300~\ang\ emission line in
our  observed spectra  is of  telluric origin.   We did  not  find any
detectable  emission  in the  \hb,  \ndo\,  and  \ndt\ line  profiles,
including the  other photospheric  lines typical to  an A7  type stars
within the noise level of  our observed spectra.  Though the \ha\ line
profiles were dominated by variable absorption components, but nominal
to significant variable emission components were also observed.  Below
we describe  in detail the  behaviour of different line  profiles.  To
compare  the  spectra  obtained  in  different  epochs  one  needs  to
normalise the spectra in a  similar fashion.  The normalisation of the
echelle spectra by fitting the continuum is tricky, because each order
displays  a  limited  wavelength  coverage  (in  this  case  \near  40
\AA). However, with  the synthetic spectra of the  star as the guiding
standard,  a better  fit to  the  continuum and  accordingly a  common
normalisation of  all spectra  can be obtained.   As we  shall discuss
below,  synthetic   spectra  are   also  essential  to   identify  the
circumstellar  component in the  line profiles  that are  dominated by
absorption components.
\subsection{Stellar parameters and synthetic spectra}
\cite{balona2002}  estimated the effective  temperature (T$_{eff}$\,),
the rotational velocity  ({\it v }sin {\it i}), and  the log {\it g}\,
(logarithm of the  surface gravity) of \snm\ in  order to generate the
synthetic spectra of  the star.  \cite{ripepi2003} estimated effective
temperature  (T$_{eff}$  \near7425-7600  K)  of  the  star  using  the
pulsation    models.    The    average   effective    temperature   of
\citeauthor{ripepi2003} matches quite well with the estimated value of
7500~K  by  \citeauthor{balona2002}.  Recent  studies  on the  stellar
parameters of  \ac{HAeBe} stars showed  that the typical value  of log
{\it  g}  is $\sim$4.0,  which  also gives  a  reasonable  fit to  the
photospheric  wings of  the \ha\  and \hb\  profiles of  emission line
\ac{HAeBe} stars.  We used the  model atmosphere from the Kurucz model
atmosphere  database\footnote{\url{http://kurucz.harvard.edu/}} for an
A7 III star with  T$_{eff}$\,=7500 K as adopted from \cite{balona2002}
and log  {\it g}\,=  4.00.  We used  {\it v}~sin{\it i}=102  km s$^-1$
\citep{balona2002}  and  a micro  turbulent  velocity  of  2 \kms\  as
suggested  by  \cite{dunkin1997}.   Detailed  synthetic  spectra  were
computed with  the interactive  data language (IDL)  interface SYNPLOT
(I.   Hubeny,   private  communication)  to   the  spectrum  synthesis
programme  SYNSPEC \cite{synspec2000},  utilising  the adopted  Kurucz
model  atmosphere.   The  linelist  provided  in SYNSPEC  is  used  to
generate the synthetic spectra.
\begin{figure}
\begin{center}
\includegraphics[height=\columnwidth,width=\columnwidth]{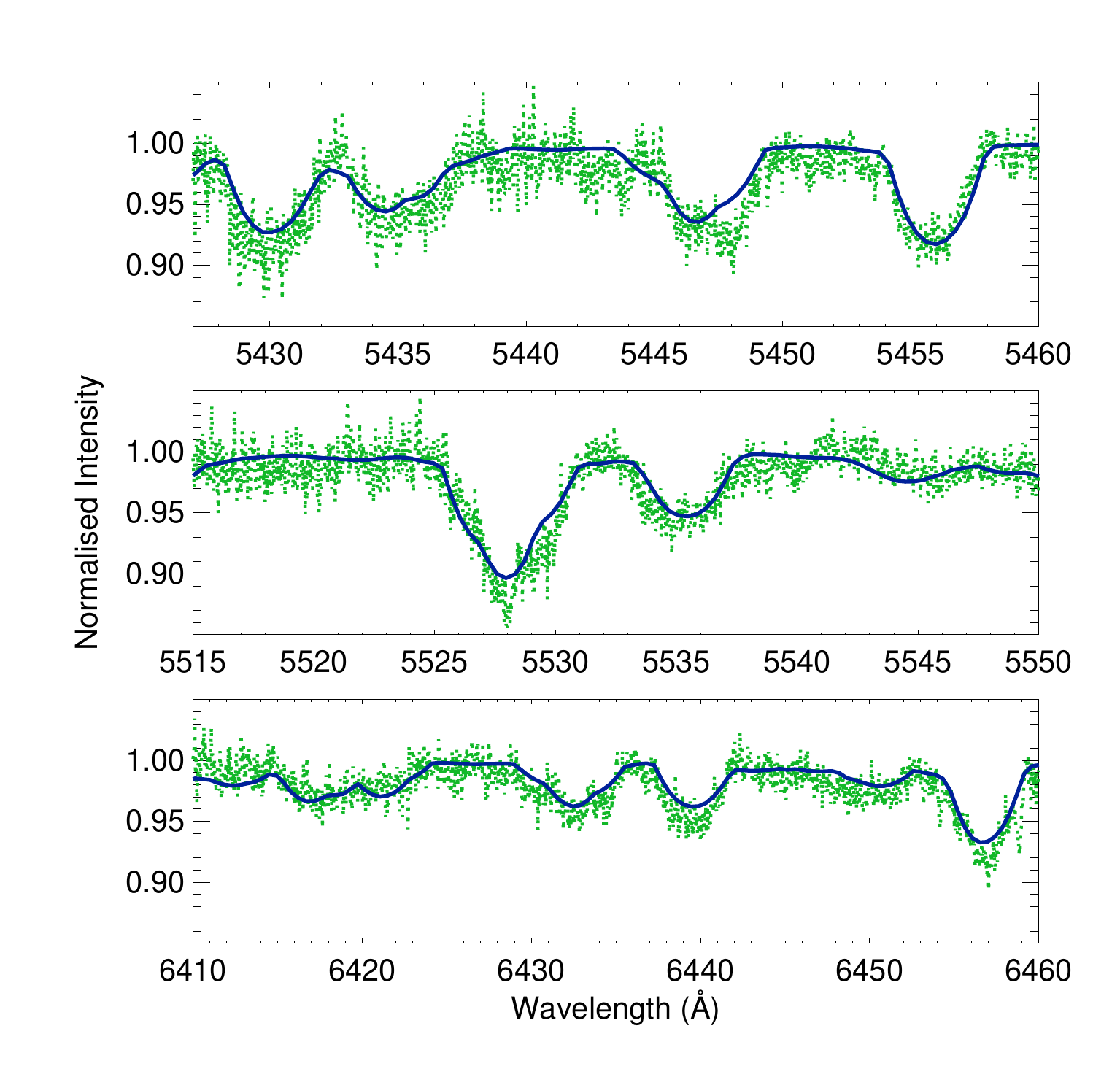}
\caption{Average    photospheric   line    profiles   of    \snm\   on
 29~December~2009 (\textit{dotted green}  line).  Synthetic spectra in
 the  same  wavelength  range  are  overplotted  (\textit{solid  blue}
 line). The prominent photospheric  lines are Fe~I 5429.696, 5434.523,
 5445.042, 5446.871, 5446.916, 5455.441 and 5455.609 in the first row,
 Sc~II 5526.790, Mg~I  5528.405 and Fe~II 5534.847 in  the second row,
 and Fe~I 6430.844, Fe~II  6432.680, Ca~I 6439.075, and Fe~II 6456.383
 in the third  row. (This figure is available  in colour in electronic
 form.)}
\label{fig_synspec}
\end{center}
\end{figure}

\begin{figure}
\begin{center}
\includegraphics[height=\columnwidth,width=\columnwidth]{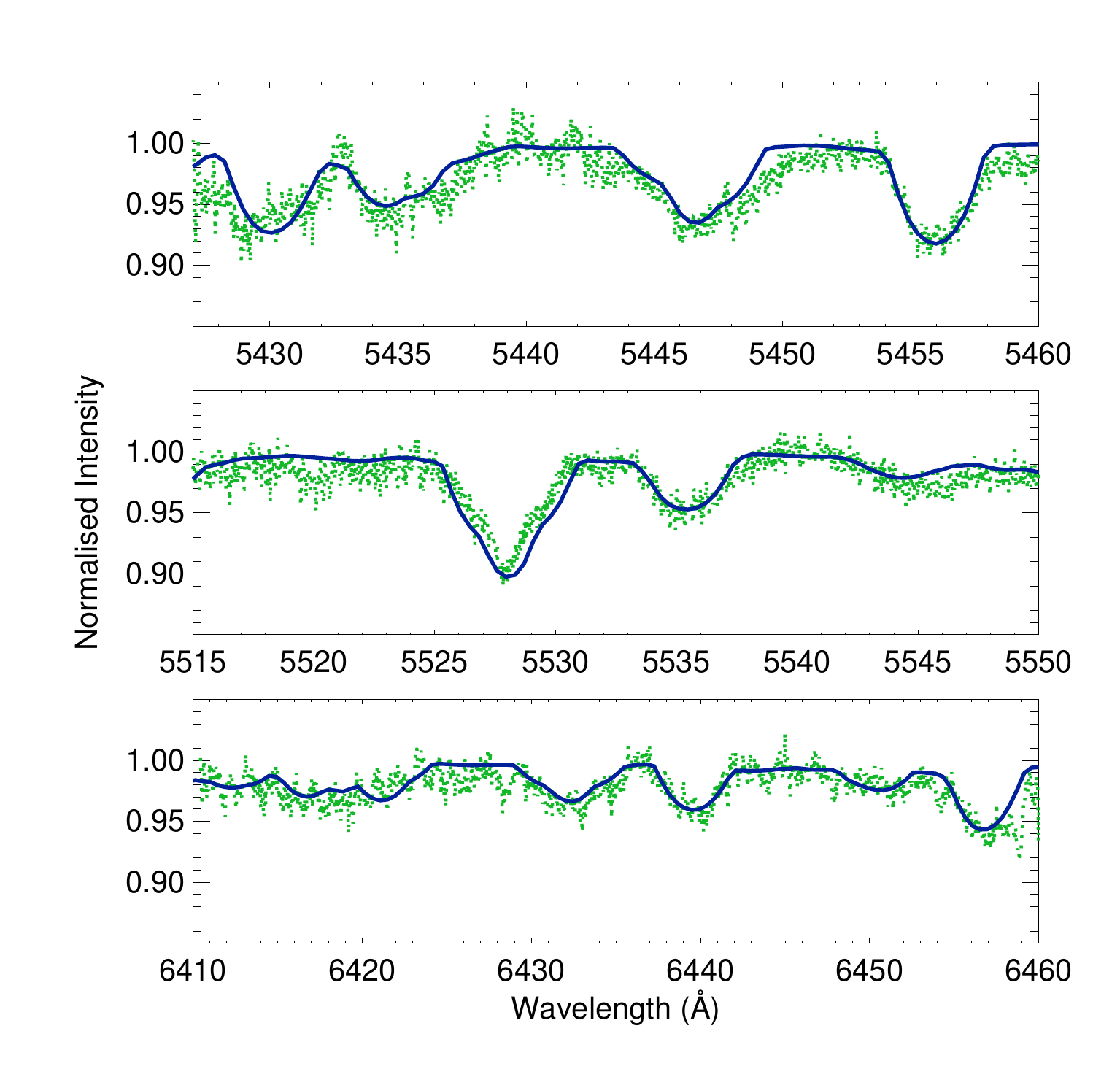}
\caption{Average  photospheric  line  profiles   of  \snm\  of  17  to
  19~January~2009 (\textit{dotted green}  line).  Synthetic spectra in
  the  same  wavelength  range  are overplotted  (\textit{solid  blue}
  line). (This figure is available in colour in electronic form.)}
\label{fig_synspec_jan}
\end{center}
\end{figure}
We  convolved the  synthetic spectra  with the  rotational  profile of
\snm\ and  the instrumental profile  of the echelle  spectrometer with
SYNPLOT                                                             and
ROTINS\footnote{\url{http://nova.astro.umd.edu/Synspec43/synspec-frames-rotin.html}}.
In Figs.~\ref{fig_synspec}  and \ref{fig_synspec_jan} we  display some
of  the representative  photospheric  line profiles  and overplot  the
convolved  synthetic spectra.  Because  the observed  photospheric and
synthetic line profiles show a good match over long wavelength ranges,
we did not attempt to  fit the photospheric lines rigorously to obtain
the stellar  parameters.  We used  the convolved synthetic  spectra to
identify unblended photospheric lines to obtain the radial velocity of
the the  star.  We used  Fe~I, Fe~II, and  Ca~I lines in  a wavelength
range   from    $\lambda$5000   to   $\lambda$6500~\ang\    [such   as
Fe~II~$\lambda$4923.827~\ang,              Fe~I~$\lambda$5001.862~\ang,
Fe~I~$\lambda$5429.696~\ang,              Fe~II~$\lambda$5534.847~\ang,
Fe~I~$\lambda$5615.644~\ang,               Ca~I~$\lambda$6102.723~\ang,
Ca~I~$\lambda$6122.217~\ang,  Ca~I~$\lambda$6439.075~\ang]  to measure
the radial velocity of the star.  We estimate the average heliocentric
radial velocity of the star as $+~11_{-3}^{+4}$~\kms, which is similar
to    the   estimated    radial   velocity    i.e.     +~13~\kms\   by
\citeauthor{balona2002}.   We synthesised  the photospheric  \ha\ line
profile following the same procedures.
\subsection{\ha\ line profiles}
\begin{figure}[t]
\begin{center}
\includegraphics[height=6.67cm,width=\columnwidth]{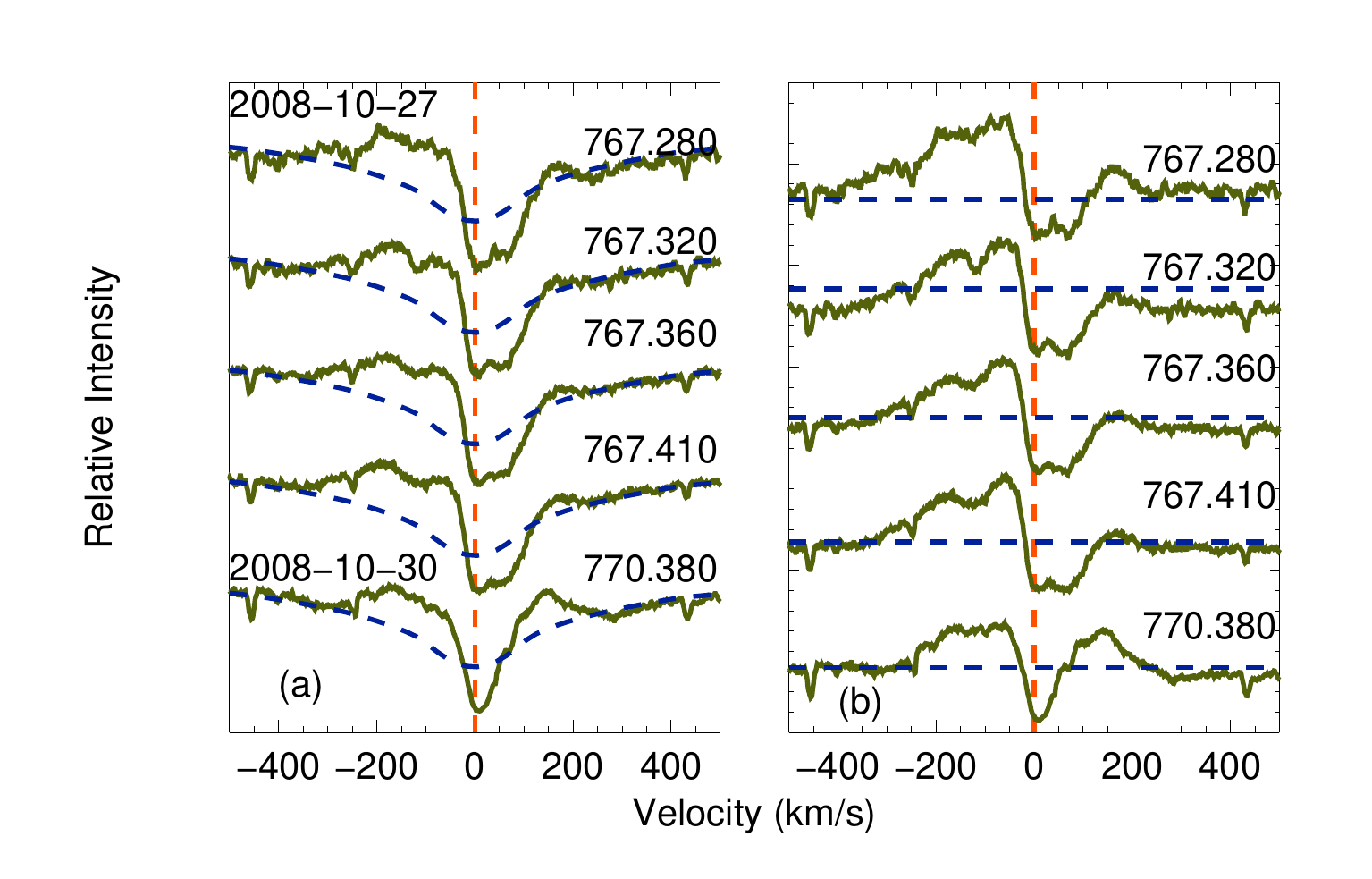}
\caption{\textbf{(a)}. \ha\ line profiles (\textit{green solid lines})
  of \snm\  on 27th  October 2008.  The  synthetic \ha\  line profiles
  (\textit{blue    dashed   lines})    are    overplotted   in    each
  spectrum. Modified Julian days  are displayed against the respective
  spectrum. The \textit{vertical red  dashed line} represents the rest
  frame velocity of the  star. \textbf{(b)}.  Residual spectra of \ha\
  line profiles on same  epochs. \textit{Horizontal blue dashed lines}
  represent the zero absorption  levels.  (This figure is available in
  colour in electronic form.)}
\label{fig_octprof}
\end{center}
\end{figure}

\begin{figure}[htbp]
\begin{center}
\includegraphics[height=13.3cm,width=\columnwidth]{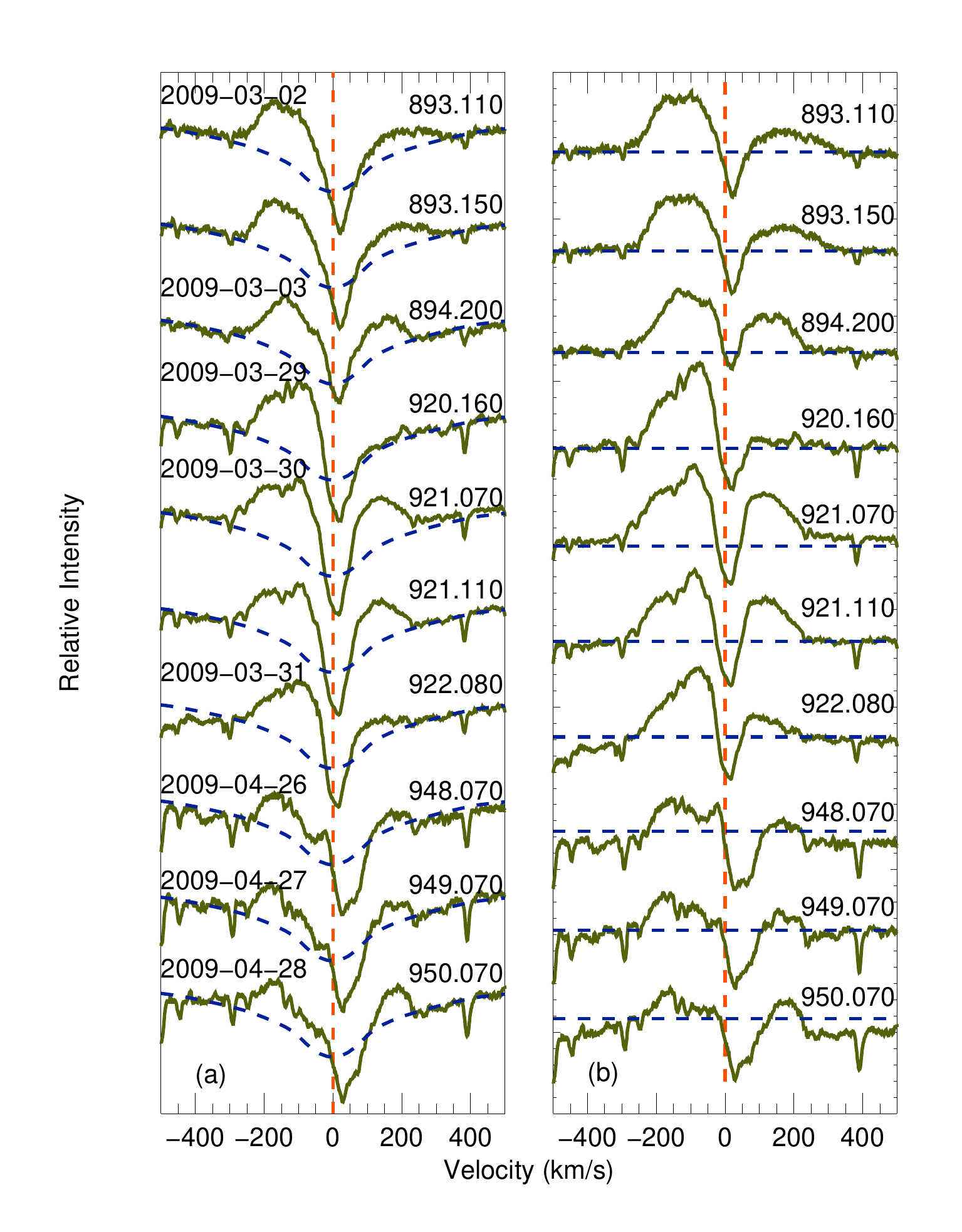}
\caption{\textbf{(a)}. \ha\ line profiles (\textit{green solid lines})
  of \snm\ of  some days in March and  April~2009. Synthetic \ha\ line
  profiles  (\textit{blue  dashed  lines})  are  overplotted  in  each
  spectrum. Modified Julian days  are displayed against the respective
  spectrum. The \textit{vertical red  dashed line} represents the rest
  frame velocity of the  star. \textbf{(b)}.  Residual spectra of \ha\
  line  profiles in  the same  epochs. \textit{Horizontal  blue dashed
  lines}  represent  the  zero  absorption levels.   (This  figure  is
  available in colour in electronic form.)}
\label{fig_marprof}
\end{center}
\end{figure}

We  display all  the \ha\  emission line  profiles in  the  rest frame
velocity  of  \snm\  and   overplot  the  photospheric  components  in
Fig~\ref{fig_octprof}(a)  to Fig~\ref{fig_marprof}(a).   We  also mark
the spectra by their date of observation and the modified heliocentric
Julian day  (MHJD) defined by 2454000--heliocentric  Julian day (HJD).
We subtract the photospheric  component from the observed spectrum and
show   the   residual    spectrum   in   Fig~\ref{fig_octprof}(b)   to
Fig~\ref{fig_marprof}(b). The  zero level in the  residual spectrum is
described as the zero  absorption level.  As discussed before, profile
variations on a timescale of days  to months are quite common to \snm.
We also  observed both blue- and red-shifted  absorption components of
various widths and  depths in several epochs. These  features are also
seen in other \ac{HAeBe} stars such as UX~Ori, BF~Ori, SV~Cep, WW~Vul,
and  XY~Per,  see   e.g.   \cite{mora2002,  mora2004}  and  references
therein.   These features  are called  transient  absorption component
(TAC) because  they are  sporadic in  nature and last  for a  few days
only.   Transient  absorption  components  can  be  divided  into  two
categories,  i.e  blue-shifted absorption  components  (BACs) and  red
shifted absorption  components (RACs).  The  chemical compositions and
kinematics of  the \ac{TACs} have  been discussed by  several studies,
e.g.   \cite{natta2000,mora2002,mora2004}.   We  discuss  the  profile
variations  in  detail  in  the  subsequent sections.   We  quote  the
velocity of the broad wings from  the residual spectra and that of the
\ac{TACs} from the observed normalised spectra.
\begin{figure}[htbp]
\begin{center}
\includegraphics[height=20cm,width=\columnwidth]{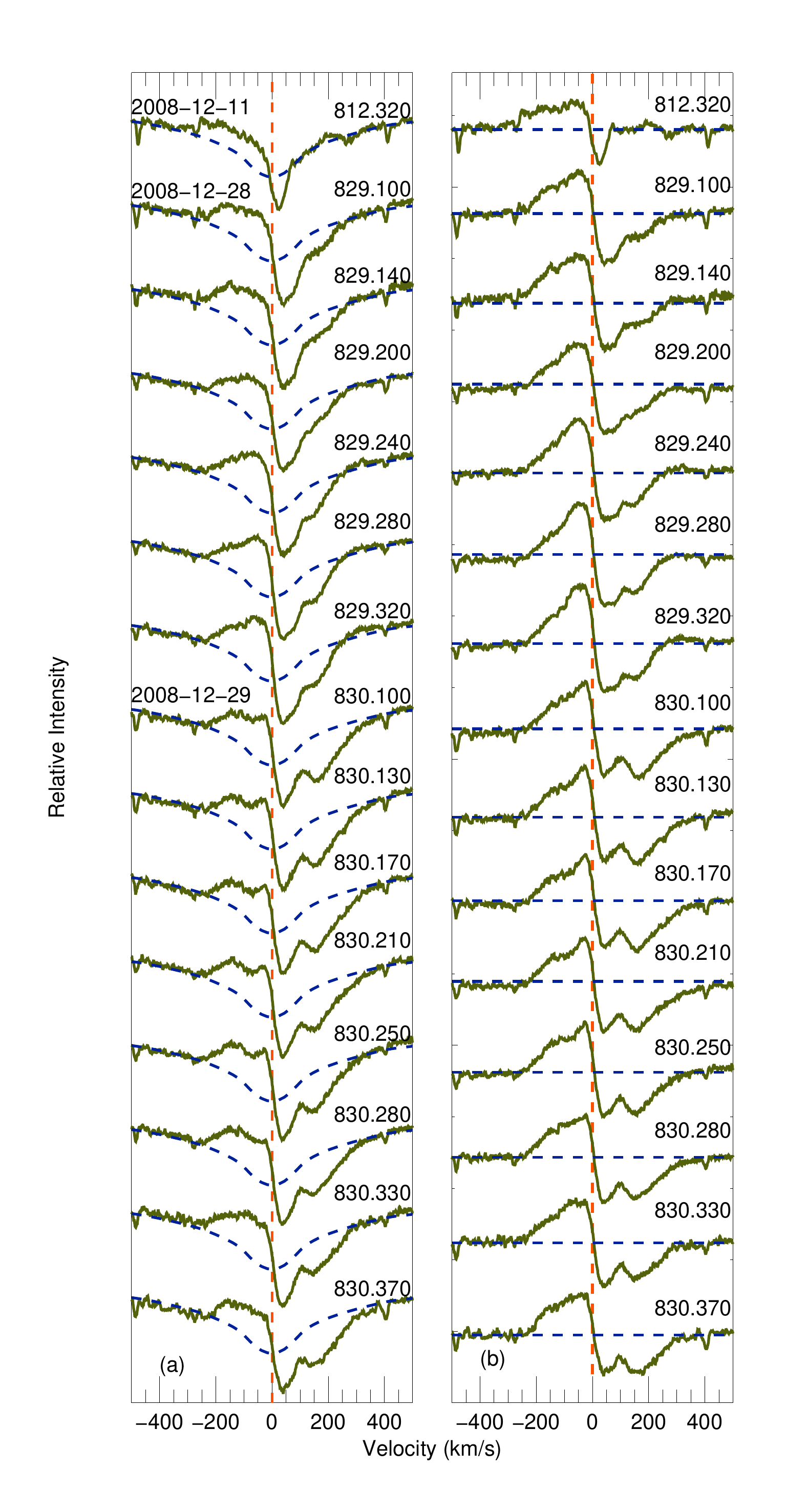}
\caption{\textbf{(a)}. \ha\ line profiles (\textit{green solid lines})
  of  \snm\  of some  days  in  December  2008.  Synthetic  \ha\  line
  profiles  (\textit{blue  dashed  lines})  are  overplotted  in  each
  spectrum. Modified Julian days  are displayed against the respective
  spectrum. The \textit{vertical red  dashed line} represents the rest
  frame velocity of the  star. \textbf{(b)}.  Residual spectra of \ha\
  line  profiles in  the same  epochs. \textit{Horizontal  blue dashed
  lines}  represent  the  zero  absorption levels.   (This  figure  is
  available in colour in electronic form.)}
\label{fig_decprof}
\end{center}
\end{figure}
\begin{figure}[htbp]
\begin{center}
\includegraphics[height=20cm,width=\columnwidth]{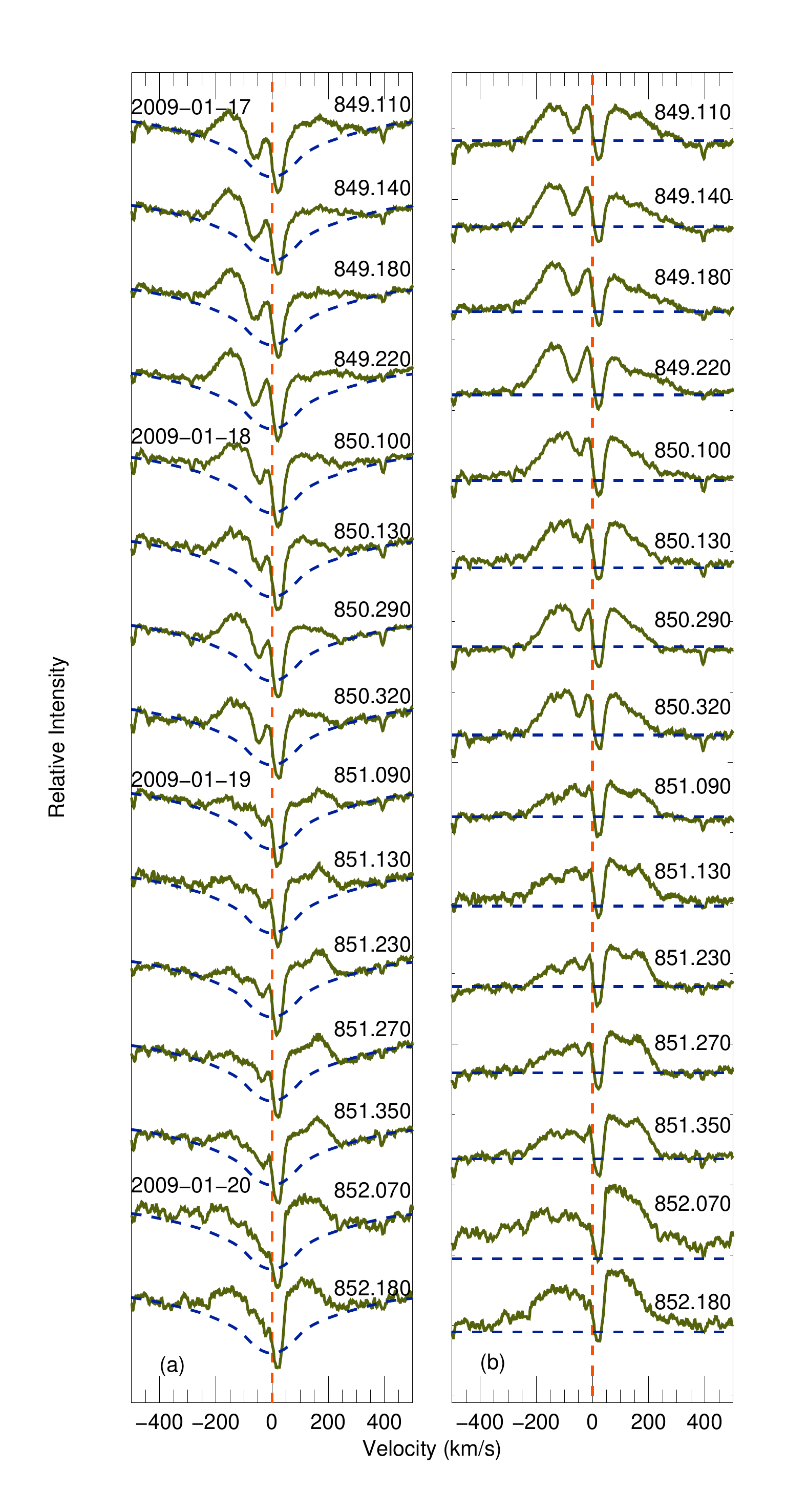}
\caption{\textbf{(a)}. \ha\ line profiles (\textit{green solid lines})
  of \snm\ of some days  in January~2009. Synthetic \ha\ line profiles
  (\textit{blue    dashed   lines})    are    overplotted   in    each
  spectrum. Modified Julian days  are displayed against the respective
  spectrum. The \textit{vertical red  dashed line} represents the rest
  frame velocity of the star.  \textbf{(b)}.  Residual spectra of \ha\
  line  profiles in  the same  epochs. \textit{Horizontal  blue dashed
  lines}  represent  the  zero  absorption levels.   (This  figure  is
  available in colour in electronic form.)}
\label{fig_janprof}
\end{center}
\end{figure}
\subsection{Line profile variations}
Previous studies of e.g.  \cite{vanancker1996}, \cite{balona2002}, and
\cite{vieira2003}  reported  that   the  average  \ha\  emission  line
profiles    of    \snm\    were    inverse   P~Cygni    types.    From
Table.~\ref{tab_obslog}  it  is  evident  that  \snm\  showed  various
profile shapes such as II-R and III-R, including inverse P~Cygni types
during our observation. Nightly variations of profiles from an average
behaviour  at a  timescale  of a  day  were also  observed in  several
epochs.  The  type II-R  profile of 2~March~2009  evolved into  a type
III-R  on 3~March~2009.   Another  incident of  a  rapid variation  in
profile  shapes  was  observed   during  29  to  31~March~2009.  Three
different types of  \ha\ profiles, namely IV-R, II-R,  and III-R, were
observed on these  consecutive nights.  The flip-over from  type II to
type III profiles, i.e the change in strength of blue- and red-shifted
emission  peaks  in  a  double-peak  profile,  was  also  reported  by
\cite{vieira2003}  in PDS~018  and PDS~024,  but on  longer timescales
that were separated  by months to year respectively.  We also observed
two prominent incidents of infall  and outflow in the \ha\ profiles on
28 to 29~December, 2008 and  17 to 20~January, 2008 respectively.  The
infall   event    was   seen   in    the   \hb\   profile    as   well
(Fig.~\ref{fig_hbnaprof}). We also checked a few photospheric metallic
lines  in  these  epochs  that  were separated  by  nearly  two  weeks
(Figs.~\ref{fig_synspec} and  \ref{fig_synspec_jan}) and did  not find
any variation in  those profiles, which also indicates  that there was
no significant variation in  the stellar photosphere during the infall
and outflow events.

\begin{figure}
\begin{center}
\includegraphics[height=6cm,width=\columnwidth]{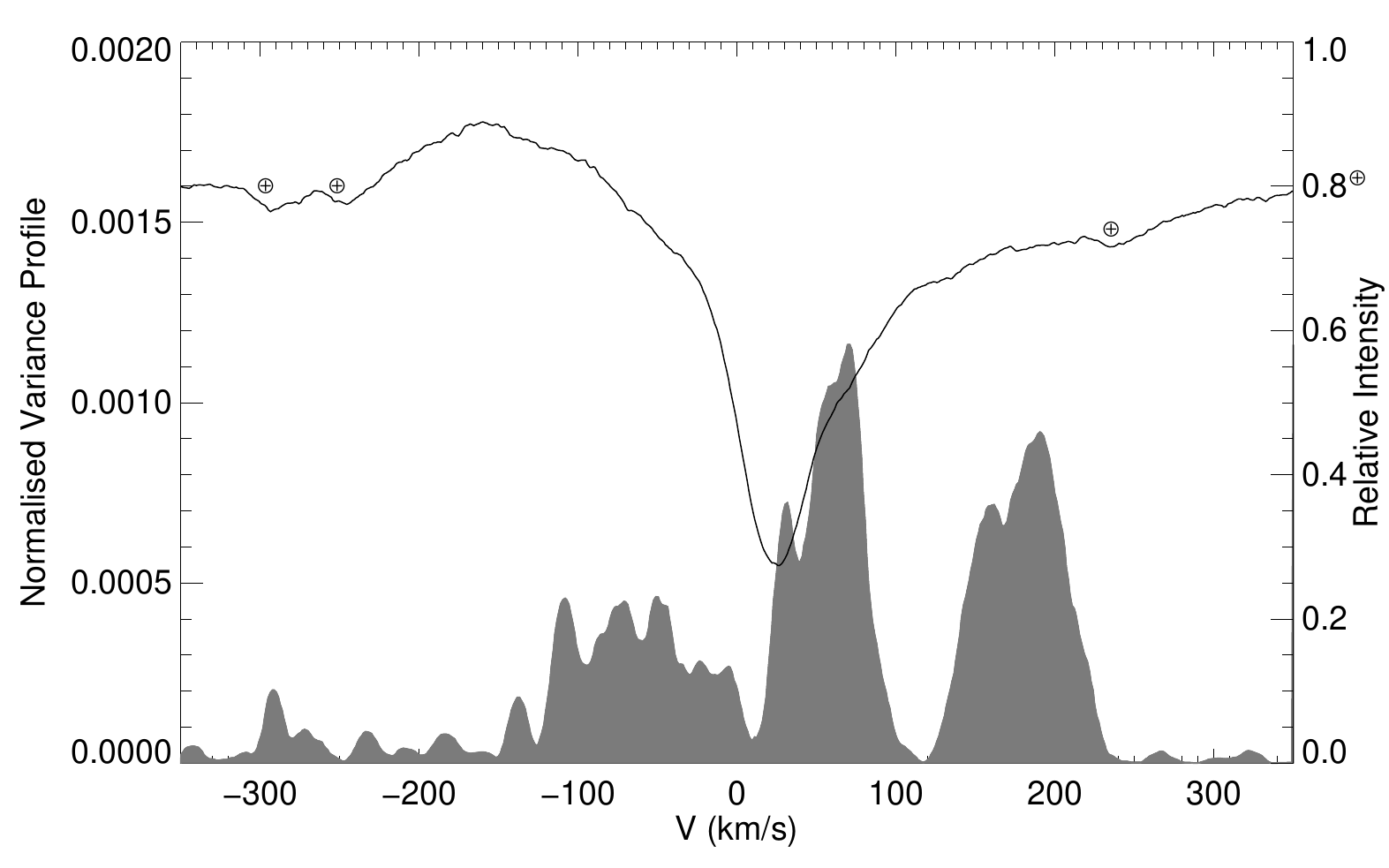}
\caption{Average \ha\  line profile (solid line)  and variance profile
  (grey shaded area) of \snm\ obtained using all  spectra.}
\label{fig_variance}
\end{center}
\end{figure}

We  calculated the  mean and  temporal variance  profile of  \ha\ line
profiles  to investigate  the characteristics  of the  variations.  We
calculated the mean profile by taking the average normalised intensity
in  each  velocity interval  of  a  given  profile over  all  observed
profiles.  We show  the average normalised \ha\ line  profile of \snm\
in  Fig.~\ref{fig_variance}.   As  suggested by  \cite{john1995a},  we
define the temporal variance profile at each velocity interval as
\begin{center}
\begin{eqnarray*}
 \hspace*{2.5cm} \sum_{v}=\frac{\sum_{i=1}^{n}(I_{v,i}-\bar{I_{v}})^{2}}{n-1},
\end{eqnarray*}
\end{center}
where n is the number of spectra, $I_{v,i}$ is normalised intensity at
a given velocity  (v) in each spectrum (i),  and $\bar{I_{v}}$ is mean
normalised  intensity  at  a  given  velocity (v)  over  all  observed
spectra.   The normalised variance  profile (shown  as shaded  area in
Fig.~\ref{fig_variance})  can  be obtained  by  dividing the  temporal
variance  profile by  the  average profile.   The normalised  variance
profile  indicates  that  the  behaviour  of  an  individual  spectrum
observed at  different epochs deviates substantially  from the average
profile. Most of  the variability in \ha\ profile is  on red side with
two clearly  distinguished features. There  are significant variations
on blue side  as well, but the profile shows a  single peak.  A couple
of outflow events were observed  at different epochs separated by days
to months,  but they seem to  have an approximate  average velocity of
-100~\kms.  We also observed simultaneous infall and outflow events of
different  durations  at different  velocities.   It  is obvious  from
Fig.~\ref{fig_variance}  that  the   infall  events  occurred  at  two
distinctly   different   velocities    of   e.g.    \near60~\kms   and
\near150~\kms.
In  emission  line  stars,  circumstellar  emission  first  fills  the
photospheric absorption  components and  the residual flux  emerges as
\textit{emission}  from the star.   If there  is strong  emission, the
reduction of emission flux to  cover up the photospheric absorption is
not  significant.  However,  for weak  emission, where  the  \ha\ line
profile is dominated by absorption rather than emission, a subtraction
of   photospheric  component   is  necessary   to  achieve   the  true
contribution  caused  by   emission.   Red-shifted  narrow  absorption
components in  Fig.~\ref{fig_octprof}(a) to Fig.~\ref{fig_marprof}(a),
which indicate  infall of matter,  are prominently visible  beyond the
photospheric  profiles.  Accordingly  the subtraction  of photospheric
profiles are  not crucial for  comparative studies of  these features.
However,   the  broad   blue-   and  red-shifted   emission/absorption
components at  the wings sometimes appeared  at the same  level as the
photospheric profiles.  To make sure that  we were not  looking at the
photospheric, but  at the circumstellar components,  we subtracted the
photospheric spectra from the  observed normalised spectra and plotted
the        profiles       in        Fig.~\ref{fig_octprof}(b)       to
Fig.~\ref{fig_marprof}(b).   The  residual  profiles  also  provide  a
better  estimate of  maximum  velocity  reached at  the  wings of  the
profiles, which  are not contaminated by  photospheric components.  We
define  the  maximum   velocity  of  the  profile  at   wings  as  the
full-width-at-zero  intensity  of   the  residual  spectra.   We  also
calculate the average and variance profiles using residual spectra.
\begin{figure}[h!]
\begin{center}
\includegraphics[height=6cm,width=\columnwidth]{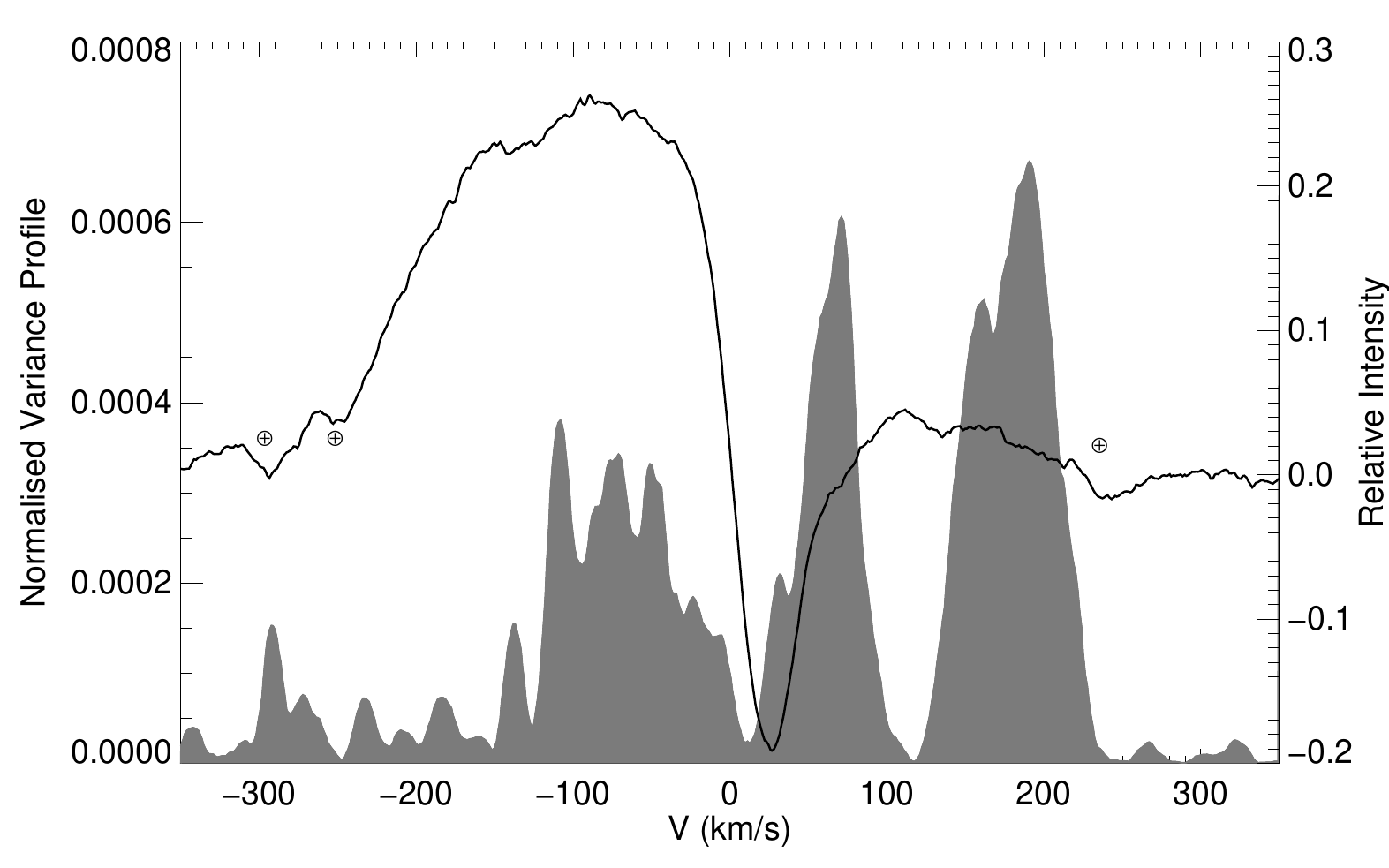}
\caption{Average \ha\  line profile (solid line)  and variance profile
  (grey  shaded  area)  of  \snm\  obtained  using  all    residual
  spectra.}
\label{fig_varianceres}
\end{center}
\end{figure}
 The  average profile  in Fig~\ref{fig_varianceres}  is  a double-peak
 profile with a strong blue  to red asymmetry, i.e.  III-R type, which
 is different  from its  counter part in  Fig~\ref{fig_variance}.  The
 red-  shifted  absorption near  the  centre  is  similar to  that  of
 Fig~\ref{fig_variance}  and  there  is  no  change  in  the  variance
 profiles  in  Figs~\ref{fig_variance}  and  \ref{fig_varianceres}  In
 Fig.~\ref{fig_varianceres}, the two peaks of the red-shifted variance
 profiles do  not coincide with  red-shifted emission peak  of average
 profile  of the residual  spectra, which  supports our  assumption of
 emission in most  epochs.  Double peak profiles are  supposed to form
 in stellar or disk winds  contrary to inverse P~Cygni profiles, which
 originate in  the infall  of materials.  We  shall discuss  below the
 effects  of   accretion,  winds,  etc.   on  the   profile  shapes  .
 Significant     changes     in     the    average     profiles     in
 Figs.~\ref{fig_variance}   and    \ref{fig_varianceres}   show   that
 photospheric profile subtraction is  necessary at least for weak \ha\
 emission line stars.
\subsection{Profile variations on 27 and 30 October, 2008}
We observed a  double-peak profile, e.g.  a type  III-R profile, on 27
October~2008 with moderate  blue-shifted and weak red-shifted emission
components      just      above      the     photospheric      spectra
(Fig.~\ref{fig_octprof}(a)).  The  blue-shifted emission component did
not  change much, but  the red-shifted  component gained  strength and
became  comparable to  its blue-shifted  counterpart, i.e.  there were
nearly  symmetric  broad emission  components  on  both  sides of  the
central minima, and thus the profile shape evolved into a type II-R on
30 October.   The maximum average  velocity of the profile  was \near\
--350~\kms\ and  \near250~\kms\ on 27 October,  and \near\ --300~\kms\
and  \near280~\kms\  on   30  October  at  the  blue   and  red  wings
respectively.  We  observed simultaneous inflow and  outflow events on
27   October.   The  central   broad  asymmetric   absorption  feature
(\near40~\kms) on 27 October was extended more towards the red region.
It  consisted  of  two  barely  resolved components  with  centres  at
\near6~\kms  and  \near50~\kms.   This  broad absorption  might  be  a
combination  of two absorptions  of nearly  similar depths  because of
both the disk and the infall of the circumstellar material.  The width
of  the central absorption  was reduced  on 30  October and  seemed to
represent  a single  absorption  component at  \near9~\kms, which  was
nearly symmetric  with the  rest frame velocity  of the star.  We also
observed a blue-shifted absorption  component on 27 October superposed
on  the blue-shifted  emission  component, which  is  considered as  a
signature of the outflow.  The full width at half maxima (FWHM) of the
outflow    increased   with    time,   but    its    depth   decreased
(Table.~\ref{tab_tacs}).
\subsection{Profile variations on 11, 28 and 29 December, 2008}
\begin{figure}[h!]
\begin{center}
\includegraphics[height=\columnwidth,width=\columnwidth]{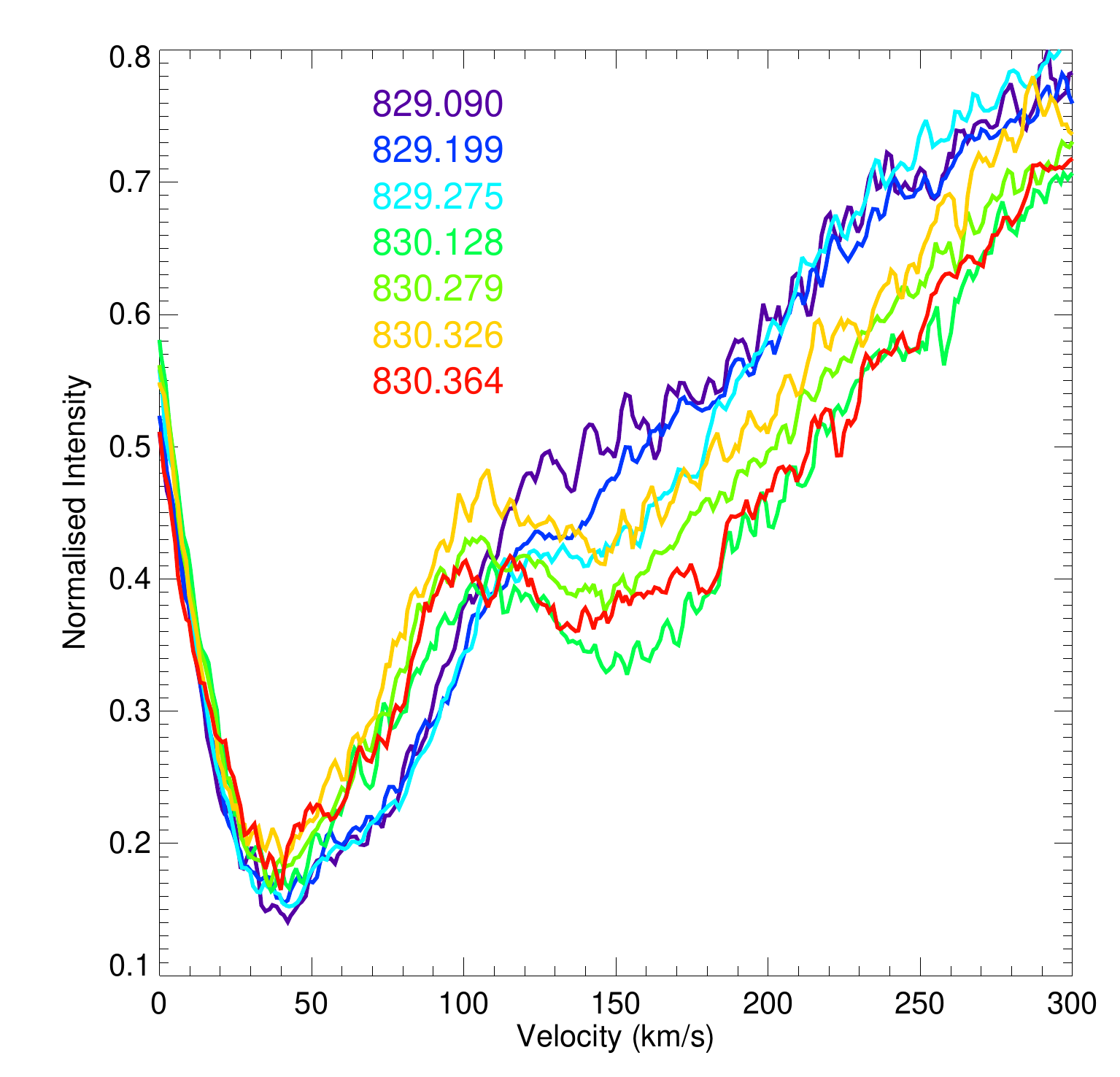}
\caption{Selected   \ha\   emission    line   profiles   on   28   and
  29~December~2008   in  the   rest  frame   velocity  of   \snm\  are
  overplotted. MHJD of the respective profiles are displayed using the
  same colour as of the profiles.}
\label{fig_decoverplot}
\end{center}
\end{figure}

The \ha\  line profile on  11~December~2008 was of an  inverse P~Cygni
type    with   a   narrow    absorption   centred    at   \near30~\kms
(Fig.~\ref{fig_decprof}).  It  was one of  the few profiles  where the
red component  almost matched  the photospheric component.   During 28
December and 29 December~2008 the  \ha\ line profiles were also of the
inverse P~Cygni type.  We observed the emergence of an infall event on
28th December and  followed its evolution for several  hours and until
the next night.  The  depth of the RAC was less than  the depth of the
central  absorption  component,  which  became red-shifted  and  wider
compared to  its counter part on 11~December.   The nightly variations
of  the maximum  velocity  of the  blue  wings were  between --250  to
--300~\kms, and that  of the red absorption wings  was 300~\kms and it
did  not  vary much  throughout  the  night.   On 29~December~2008  we
observed both prominent infall  and weak outflow events simultaneously
(Fig.~\ref{fig_decprof}(a)  MHJD~830.100 to  830.280).   The RAC  that
emerged on 28~December became deeper and its centre was separated from
the central  absorption component, as a  result of which  the width of
the central absorption component reduced (Fig.~\ref{fig_decoverplot}).
The maximum velocity at the blue wings did not change, but the average
maximum  velocity  at  the  red  wings  shifted  to  \near380~\kms  on
29~December. The  depth of the RAC varied  throughout the observations
on 28 and 29~December.
 We   show    the   emergence   and   evolution   of    the   RAC   in
 Fig.~\ref{fig_decoverplot}.  The velocity  of the RAC was decelerated
 within a few  hours from its starting point  and then became constant
 at MHJD~830.21, where it remained until the end of the observation on
 29~December (Table.~\ref{tab_tacs}).   All observed line  profiles of
 28  and 29 December  showed a  blue-shifted emission  component whose
 strength increased after MHJD~829.1.   We detected the outflow on top
 of the  blue-shifted emission at  MJHD~830.100 and could  identify it
 clearly before MHJD~830.330.  It  gradually moved towards the central
 absorption  features so  the  probable reason  for its  non-detection
 after MHJD~830.330 might be because the  BAC came in the same line of
 sight of the central absorption.
\subsection{Profile variations on 17 to 20 January, 2009}
The \ha\ emission line  profiles observed during 17 to 20~January~2009
are  mostly   dominated  by   both  blue-  and   red-shifted  emission
components,    i.e.    mostly    Type~II    and   Type~III    profiles
(Fig.~\ref{fig_janprof}).   The strength  of both  emission components
varied during our  observation, e.g.  one of the  component was always
weaker  compared to the  other component  and thus  the shapes  of the
profiles also  changed from one  type to another. The  average maximum
velocities at the wings were \near\ --250~\kms\ and \near 250~\kms\ in
the  blue and  red  sides  respectively.  We  also  detected a  strong
outflow,  which was  clearly detectable  from 17  to 19  January.  The
central narrow absorption component at \near20~\kms\ remained constant
throughout the observations.
\begin{figure}[h!]
\begin{center}
\includegraphics[height=6cm,width=\columnwidth]{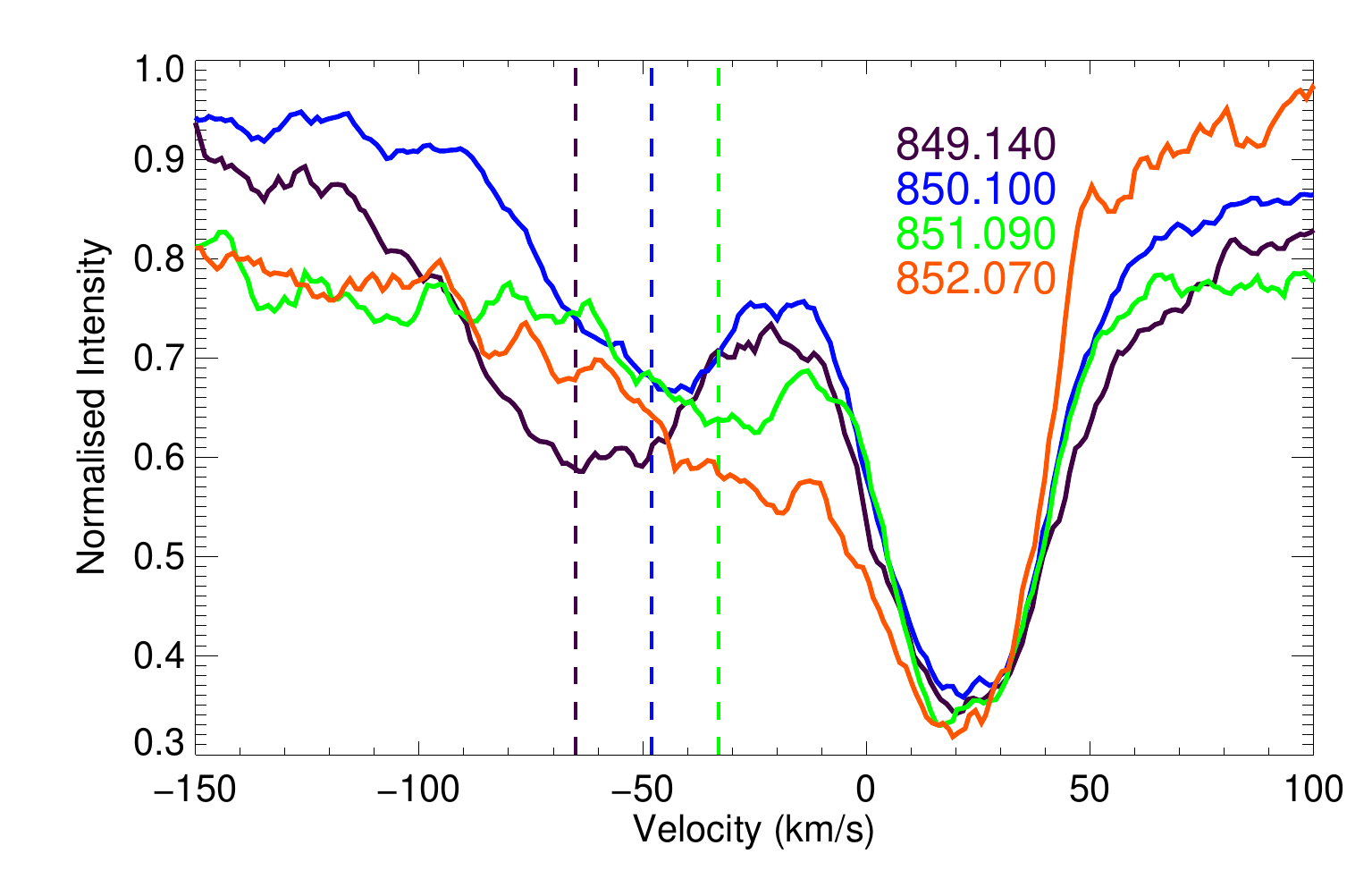}
\caption{Selected \ha\ emission line profiles on 17 to 20~January~2009
  in the  rest frame  velocity of \snm\  are overplotted. MHJD  of the
  respective profiles  are displayed using  the same colour as  of the
  profiles. Vertical  dashed lines of different  colours represent the
  centres   of  the   BACs  estimated   by  eye   on   the  respective
  MHJDs.  Red-shifted  emission increased  on  MHJD~852.070 after  the
  disappearance of the outflow event.}
\label{fig_janoverplot}
\end{center}
\end{figure}
We    display    the    time    evolution   of    the    outflow    in
Fig.~\ref{fig_janoverplot}. The  outflow gradually became  wider as it
continuously moved towards the central absorption. It is also worth to
mention that  the width of the  central absorption was  at its minimum
among our  observed spectra.  The  non-detection of the outflow  on 20
January at MHJD~852.070 was  most probably caused by the superposition
of its  line of sight with  the central absorption.  As  a result, the
central  absorption  became  wider  and asymmetric,  e.g.   it  became
elongated toward  blue side and  the red-shifted emission  reached its
maximum strength compared  to the other spectra observed  during 17 to
19 January, 2009.
\subsection{Profile variations in March and April, 2009}
Most of  the profiles we observed  on several days in  March and April
were  double-peaked profiles  (Fig.~\ref{fig_marprof}(a)).   On 2  and
3~March~2009,  there was  a change  in the  slope of  the blue-shifted
emission. The  red-shifted emission gradually gained  strength and the
profile evolved  from type  III-R to type  II-R.  The  average maximum
velocity at the blue wing  was \near\ --~250~\kms\ and \near 300~\kms\
at  the  red  wing.  We  did  not find  any  red-shifted  emission  on
29~March~2009  (MHJD~920.160).   The  profile  showed  a  blue-shifted
emission and  the central  absorption component. On  the next  day, we
observed a strong red-shifted emission comparable to the blue one, and
thus  the profile  was  changed  into a  type  II-R.  The  red-shifted
emission was again considerably reduced  on 31 March and we observed a
type III-R profile. We detected a blue-shifted absorption component at
\near71~\kms\  on  26~April~2009.   The absorption  component  shifted
towards  the deep  minima  on  27~April.  There  was  no signature  of
outflow in  the spectra on  28~April. The red-shifted  emission became
gradually stronger  from 26 to  28~April and became comparable  to the
blue-shifted emission. These changes  are similar to those observed on
28  and  29~December~2008  (Fig.~\ref{fig_decprof}), but  the  central
absorption features  were different,  e.g.  in April~2009  the central
absorption component was board and asymmetric and extended towards the
red  region  compared  to  the  narrow and  nearly  symmetric  central
absorption in December~2008.
\subsection{\hb,  \ndo, and \ndt\   profiles}
We  obtained  simultaneous  spectra  containing  the  \hb,  \ndo,  and
\ndt\ profiles along with \ha\ emission line profiles.
\begin{figure}[h!]
\begin{center}
\includegraphics[height=9cm,width=\columnwidth]{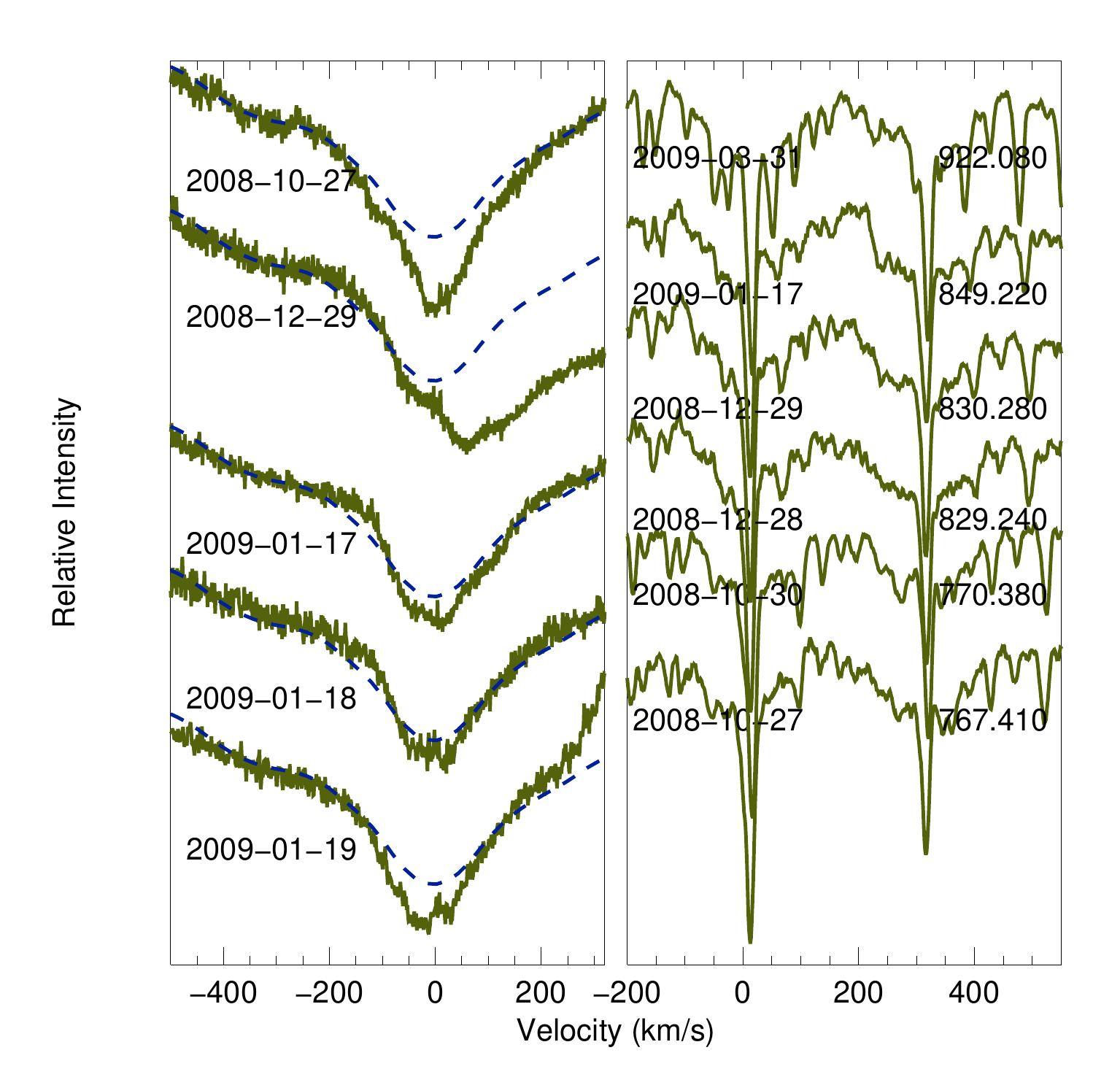}
\caption{\textbf{(a)} Nightly average \hb\  profiles of \snm\ on several
  days. \textbf{(b) } Sample \ndo \,  and \ndt \, profiles of \snm\ on
  a few epochs. Velocities are in the rest frame of \snm.}
\label{fig_hbnaprof}
\end{center}
\end{figure}
Due  to the  poor  signal-to-noise  ratio, we  could  not analyse  the
individual  \hb\  profiles.   Instead,  we plot  the  nightly  average
profiles of \hb\ in the rest  frame velocity of \snm\ and overplot the
synthetic spectra in Fig.~\ref{fig_hbnaprof}(a). The shape of the \hb\
profiles also showed  variation over the timescales of  month.  We did
not find any  emission components in the \hb\  profiles.  However, the
prominent infall event  occurred during 28 and 29~December,  as can be
easily  seen also  in the  \hb\ line  profiles.  The  \ndo\  and \ndt\
profiles  displayed in Fig.~\ref{fig_hbnaprof}(b)  consist of  a broad
and  a narrow  absorption component.   We did  not find  any elongated
absorption in  the broad line  profiles.  We measured the  velocity of
the narrow component and found  that it was showing an almost constant
velocity  of  \near\  +22~\kms,  i.e.   red-shifted  by  \near11~\kms,
compared to the star.
\section{Discussion}
\label{section:diss}
\ac{HAeBe} stars show photometric  variabilities on timescales of days
to  years.    \cite{vanancker1996}  analysed  the   long-term  optical
photometric behaviour of \snm\ and  found that the star showed a large
variation  of $>$2 mag  before 1985.   \citeauthor{vanancker1996} also
reported that the visual magnitude of \snm\ varied by 0.16 mag between
1985 to 1990,  i.e. over timescale of years. During  1990 to 1995, the
visual  magnitude  of the  star  became  almost  constant without  any
significant   variation.   \cite{balona2002}   obtained   the  optical
magnitudes   (e.g.   V$\sim$8.92,  $B-V\sim$0.37,   $V-R\sim$0.22  and
$V-I\sim$0.46 mag)  over timescales  of minutes to  weeks and  did not
find any  significant variation within 0.05 mag.   We obtained optical
\textit{BVRI} photometry of the star on 11 November, 2009 and obtained
the   magnitudes   V$\sim$8.89,   $B-V\sim$0.38,  $V-R\sim$0.19,   and
$V-I\sim$0.41  mag,  which indicates  that  there  are no  significant
changes in the long-term photometric behaviour of \snm. However, it is
also noteworthy that our photometric observation did not coincide with
the  spectroscopic  observation.   Unlike  the  variations  in  visual
magnitudes, which  reflect the variable  stellar photosphere, \ac{NIR}
variability can  account for changes  in both the  stellar photosphere
and  the circumstellar  environments.  \cite{balona2002}  compared the
\ac{NIR}  magnitudes of  \snm\ and  found that  $H-K$ excess  of \snm\
(which  serves as  a proxy  for  the inner  disk materials)  decreased
continuously over the period  from 1982 to 2002.  \cite{vanancker1996}
and   \cite{balona2002}  suggested   that   the  systematic   \ac{NIR}
variability  of \snm\ is  caused by  the clearing  of dust  around the
star,  which  indicates a  variable  circumstellar environment  around
\snm.

Our high-resolution  spectroscopic observations revealed  the variable
circumstellar  environment and  presences  of TACs  in  the weak  \ha\
emission line of the Herbig~Ae  star \snm.  As mentioned earlier, TACs
in higher Balmer lines and in  a few metallic lines have been detected
in  Herbig  Ae stars  such  as  BF~Orionis, SV~Cephei,  WW~Vulpeculae,
XY~Persei, and UX~Orionis \citep{mora2002,mora2004}.  All these Herbig
Ae stars show prominent  \ha\ emission profiles with equivalent widths
ranging  from -2.3~\ang\  (for UX~Ori  ) to  -30.0~\ang\  (for WW~Vul)
\citep{manoj06}.   These  stars  also  show  optical  photometric  and
polarimetric     variabilities    over     timescales     of    months
\citep{oudmaijer2001}.  XY~Per,  UX~Ori, BF~Ori, and  WW~Vul also show
NIR  variability  over timescales  of  months \citep{eiroa2001}.   The
characteristics of \snm\ do not match the trend of the Herbig Ae stars
mentioned   above,  because  \snm\   shows  nearly   constant  optical
magnitudes  over  years  and  a  low \ha\  emission  equivalent  width
($\leq$1~\ang).   Moreover,   \snm\   is  also   comparatively   older
($\sim$6.5~Myr) than the mentioned group  of Herbig Ae stars with ages
ranging  from 2.5  to  5.2~Myr \citep{manoj06,montesinos2009}.   These
differences make \snm\ an  interesting target to study the interaction
of intermediate PMS stars (contracting towards the main sequence) with
their circumstellar environment.

\snm\ is known  as an isolated \ac{HAeBe} star  without any associated
  nebulosity     in      its     immediate     vicinity.      However,
  \citep{hernandez2005} suggested  its association with  the Ori~OB1bc
  region located at  a distance of 400 to  450~pc.  The estimated ages
  of  \snm\  available  in  the  literature vary  from  1  to  6.5~Myr
  \citep{vanancker1996,ripepi2003}.   The main sources  of uncertainty
  to estimate the age from the HR diagram is the distance to the star.
  The lower limit of  the distance is \near210~pc \citep{hipnew07} and
  the upper limit  is \near450~pc, which comes from  the assumption of
  its     association    with     the     \textit{Ori~OB1bc}    region
  \citep{hernandez2005}.   \cite{ripepi2003}  analysed  the  pulsating
  behaviour  of \snm\  and  suggested an  intermediate  value for  the
  distance  between the  upper  and  lower limit,  which  leads to  an
  estimate  of  the  age  of  \near6.5~Myr.  The  \ha\  emission  from
  \ac{HAeBe}s, which  is a  proxy for active  accretion, are  found to
  decrease significantly beyond  3~Myr e.g. \cite{manoj06}.  There are
  some \ac{HAeBe} stars older  than $\sim$~6~Myr with significant \ha\
  emission \citep{manoj06, montesinos2009}.  However, the proposed age
  of \snm\ i.e.  6.5~Myr seems  to be consistent with the common trend
  of  weak \ha\  emission from  stars  with ages  $>$~3~Myr.  We  also
  consider  that  the  mass  of  \snm\ is  2~\msun\  as  estimated  by
  \cite{ripepi2003}.
\subsection{Kinematics of \ac{TACs}}
\begin{figure}[h!]
\begin{center}
\includegraphics[height=10cm,width=\columnwidth]{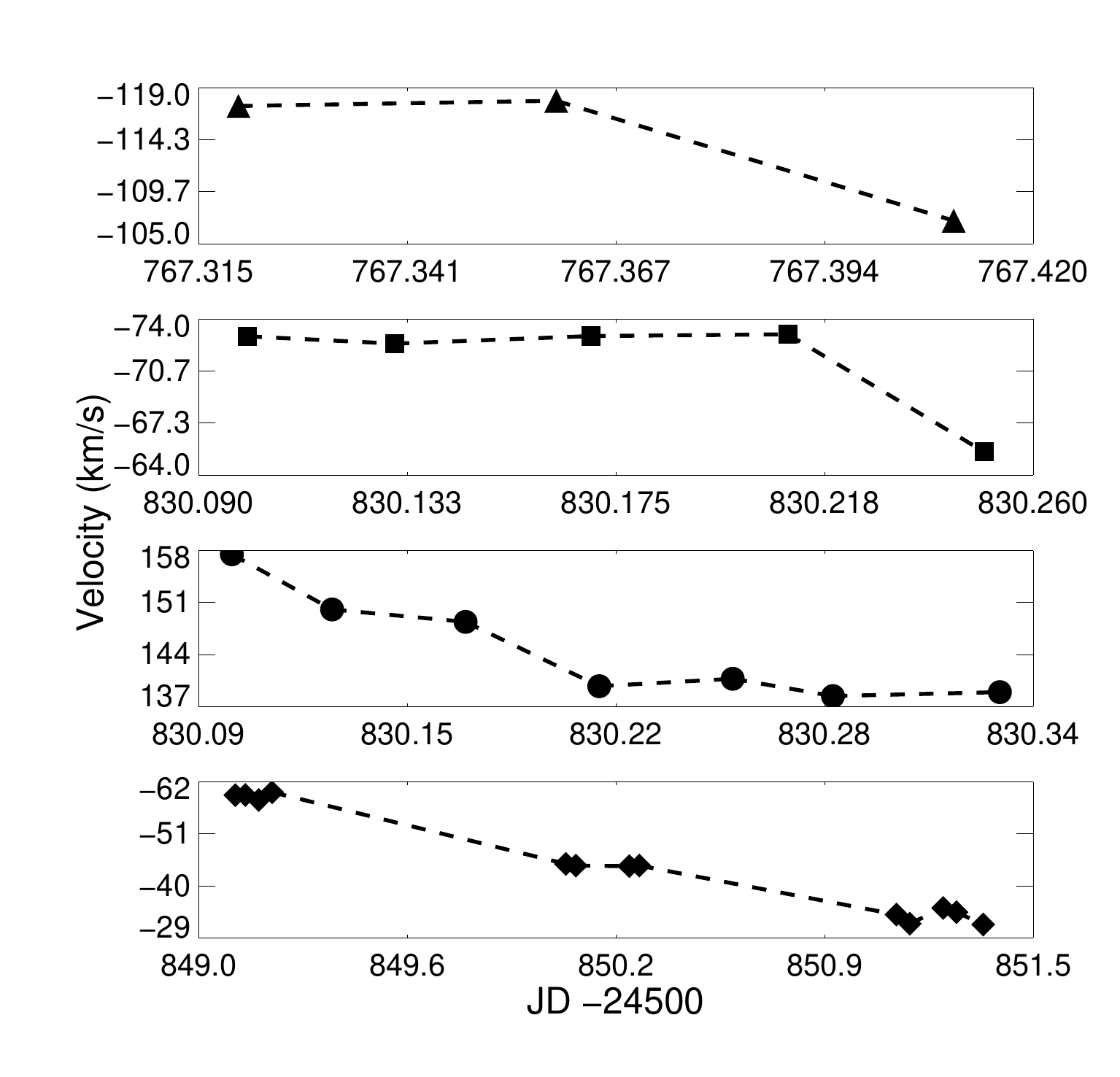}
\caption{Time evolution of the \ac{TACs}  as they appeared in the \ha\
  line  profiles.  \textit{Triangles} represent  the outflow  event on
  27~October~2008, \textit{squares} and \textit{circles} represent the
  outflow  and  infall events  respectively  on 29~December~2008,  and
  \textit{diamonds}   represent  the  outflow   event  during   17  to
  19~January~2009.}
\label{fig_tacsvel}
\end{center}
\end{figure}

Blue-shifted absorption  components (BACs) and  red-shifted absorption
components  (RACs) detected in  \ha\ emission  line profiles  of \snm\
showed variation  in the  velocity centres, depths  and widths  of the
profiles.   We   show   the   evolution   of  a   RAC   and   BAC   in
Figs.~\ref{fig_decoverplot} and \ref{fig_janoverplot} respectively. We
tried to  reproduce the observed  \ha\ profiles by fitting  a multiple
number  of Gaussian  component but  we  also noticed  that the  fitted
Gaussians  were not  unique.   When  we tried  to  fit some  profiles,
e.g. the observed  profiles of 29 to 31~March, 2009,  we found that we
got  a  good  fit  when  we  fitted  two  separate  Gaussians  to  the
blue-shifted and red-shifted emission  components rather than a single
Gaussian  for  the whole  emission  component.  The  fitting of  these
complicated  and  variable profiles  for  all  observed  spectra in  a
consistent  way demands a  complete understanding  of the  geometry of
line-forming  regions  and  their  distribution with  respect  to  the
central star, which is beyond  the scope of this work.  Accordingly we
measured  the velocity  centres and  depths of  the TACs'  profiles by
fitting  Gaussian profiles to  the individual  TACs of  the normalised
spectra over the  baseline of unity (obtained by  the normalisation of
the observed spectra by synthetic photospheric spectra).  In this way,
we  estimated  the  velocity  centres  and depths  of  the  individual
component confidently, however, we could  not estimate the FWHM of the
individual TACs, which requires the  fitting of the entire profile. As
a result  we discuss the kinematics  of the observed TACs  and not the
physical conditions  of the infalling and  outgoing materials, because
that partly  depends on  the values of  the FWHM  as well. It  is also
worth  to mention that  the detection  and analysis  of TACs  in other
Balmer  lines  such  as   \hb,  $H\gamma$  etc.   provides  additional
information.   Because of the  poor signal-to-noise  ratio at  \hb\ we
could not  detect the  TACs as \ha\  profiles. However,  good temporal
coverage  of   the  spectroscopic  observations  helped   to  make  an
unambiguous detection of the TACs  in the \ha\ line profiles, and thus
the kinematics of the TACs  estimated from \ha\ profiles only are also
highly reliable.
\begin{table}[h]
\caption{Transient Absorption Components appeared in \ha\ line profiles of \snm}             
\label{tab_tacs}     
\centering 
\begin{tabular}{lccccr}
\hline\hline
Date & HJD & Cent. & Depth & Cent. & Depth\\
yy-mm-dd&(-2,454,000)  & \kms &  & \kms &  \\
 \hline
 2008-10-27 &   767.32 &    -117.36 &      0.76 &         ... &       ... \\
 2008-10-27 &   767.36 &    -117.83 &      0.81 &         ... &       ... \\
 2008-10-27 &   767.41 &    -107.07 &      0.84 &         ... &       ... \\
 2008-12-29 &   830.10 &     157.50 &      0.34 &      -72.90 &      0.78 \\
 2008-12-29 &   830.13 &     150.07 &      0.34 &      -72.43 &      0.77 \\
 2008-12-29 &   830.17 &     148.40 &      0.33 &      -72.92 &      0.75 \\
 2008-12-29 &   830.21 &     139.74 &      0.37 &      -73.03 &      0.78 \\
 2008-12-29 &   830.25 &     140.74 &      0.38 &      -65.49 &      0.78 \\
 2008-12-29 &   830.28 &     138.41 &      0.39 &         ... &       ... \\
 2008-12-29 &   830.33 &     138.94 &      0.42 &         ... &       ... \\
 2009-01-17 &   849.11 &     -59.20 &      0.60 & ...      & ... \\
 2009-01-17 &   849.14 &     -59.24 &      0.61 & ...      & ... \\
 2009-01-17 &   849.18 &     -58.20 &      0.64 & ...      & ... \\
 2009-01-17 &   849.22 &     -59.82 &      0.62 & ...      & ... \\
 2009-01-18 &   850.10 &     -44.66 &      0.68 & ...      & ... \\
 2009-01-18 &   850.13 &     -44.30 &      0.65 & ...      & ... \\
 2009-01-18 &   850.29 &     -44.18 &      0.63 & ...      & ... \\
 2009-01-18 &   850.32 &     -44.30 &      0.62 & ...      & ... \\
 2009-01-19 &   851.09 &     -33.94 &      0.63 & ...      & ... \\
 2009-01-19 &   851.13 &     -32.02 &      0.65 & ...      & ... \\
 2009-01-19 &   851.23 &     -35.34 &      0.61 & ...      & ... \\
 2009-01-19 &   851.27 &     -34.44 &      0.60 &  ...       & ...   \\
 2009-01-19 &   851.35 &     -31.84 &      0.59 &-103.79       &0.73  \\
 2009-04-26 &   948.07 &     -55.88 &      0.61 &         ... &       ... \\
\hline  
\end{tabular}
\end{table}
We  list the velocity  centres and  depths of  the \ac{TACs}  from the
normalised continuum level in  Table~\ref{tab_tacs}.  We plot the time
evolution  of the velocities  of \ac{TACs}  in Fig~.\ref{fig_tacsvel}.
All detected  \ac{TACs} in the \ha\ line  profiles showed deceleration
of the gas at various rates.   The rate of deceleration of the outflow
event  on 27  October~2008 was  \near2~\ms,  for the  infall event  on
29~December~2008  it  was  \near1.7~\ms,  for  outflow  event  it  was
\near2.2~\ms,   and  for   the  outflow   event  during   the   17  to
19~January~2009  it was fraction  of~\ms.  Neither  the magnetospheric
accretion model  nor the wind  model provide any  detailed explanation
about  the deceleration  of the  TACs.   A detailed  calculation of  a
time-dependent  model  considering   accretion  and  wind  models  and
including  the observed results  is very  important to  understand the
variable star-disk interactions.
Broad red-shifted absorption wings  at high velocities are supposed to
originate  from the  free fall  of the  surrounding material  onto the
star. The  cool outflowing gas  is supposed to originate  farther away
from the star than  the accretion-dominated emission, but still within
about  \near10~\rsun \citep{edwards2003}.  We found  that  the average
maximum  velocity of the  red-shifted emission  wings was  \near250 to
\near300~\kms.   We detected an  enhanced red-shifted  absorption wing
extending  up to  \near380~\kms  on 29~December,  2009.  Assuming  the
free-fall  velocity condition, we  estimate the  extent of  the region
involved in  the accretion processes as  \near5~\rsun, which indicates
that   the   changes   are   taking   place   within   a   radius   of
\near0.02~AU. \cite{monnier2002}  studied the dust  sublimation radius
of the \ac{PMS} stars with  a stellar radiation field.  They discussed
the  relation  between the  size  of the  inner  dusty  disks and  the
luminosity  of the  central stars  in  the context  of different  disk
models including the  paradigm of both optically thick  and thin inner
disks.  Considering  the uncertainties such  as the distance  and thus
the  luminosity, and  the model  for optically  thin/thick  disks, the
typical  dust sublimation radius  of \snm\  would be  in the  range of
\near~0.07  to .2~AU.   A comparison  between  the region  of the  gas
accretion and the dust sublimation radius hint towards the presence of
gaseous  material   in  the  inner  disk  cavity   of  \snm.   Further
investigation of the inner disk  properties of PMS stars such as \snm\
will be  helpful to understand the disk-dissipation  processes and the
associated timescales.
\subsection{Dust mass estimates using IRAS fluxes}
The  detection  of red-shifted  absorption  wings  that extend  beyond
300~\kms\  supports  a  scenario   of  the  accretion  of  surrounding
materials  onto the  star.  However,  as \snm\  also  shows weak  \ha\
emission,  so  it  will  be  interesting to  investigate  whether  the
left-over  gas  in  the  circumstellar  environment  of  the  star  is
sufficient to continue  the accretion.  We estimate the  dust mass and
thus  the  gas mass  in  the circumstellar  disk  of  \snm\ using  the
\ac{IRAS} fluxes.

Dust  grains  present  in  the circumstellar  environment  absorb  and
scatter  varying fractions of  the incident  photons ($\lambda  \leq 1
\micron$)  and re-emit  part of  the absorbed  energy in  the infrared
band.  Far infrared emission properties of the dust grains can be used
to derive  estimates on the  amount of dust  of the disks  of \ac{PMS}
stars, assuming  the emission at these wavelengths  is optically thin.
As suggested by \cite{hildebrand1983},  the disk dust mass M$_{d}$ can
be computed using the relation
\begin{eqnarray*}
M_d  = \frac{4\pi}{3}\, \frac{a  \,\rho_d\,}{Q_\nu(T_d,a)}\, \frac{d^2
  F_\nu(T_d)}{ \pi B_\nu (T_d)},
\end{eqnarray*}
where $a, \, \rho_d$ and $d$ are the grain radius, specific grain mass
density,  and  distance  to  the  source,  respectively.   $F_\nu,  \,
Q_\nu(T_d,a)  \,$, and $B_\nu  (T_d)$ are  the observed  flux density,
grain emissivity and  the Planck function of temperature  $\rm T_d$ at
frequency $\nu$.  \cite{vanancker1996} analysed the \ac{SED} of \snm\,
and estimated  that the dust  temperature corresponds to  the \ac{FIR}
\ac{IRAS}  fluxes  as 80~K.  The  \ac{FIR}  \ac{SED}  of \snm\,  peaks
between 25  and 60  \micron\, and drops  gradually between 60  and 100
\micron, which  suggests that  most of the  emission arises  beyond 25
\micron.  In that case the calculation of dust grain temperature using
60  and 100  \micron\,  flux densities  \citep{helou1988}  would be  a
better  estimate.   Using  the  following  relation,  $\rm  T_d  \,  =
49\left(\frac{S_{60}}{S_{100}}\right)  ^{0.4}$  \citep{young1998},  we
estimated the dust grain temperature  as 53~K. We adopted $\rho_d = 3$
g\,$cm^{-3}$, $a = 0.1\,  \micron\ $ and $\frac{4a\,\rho_d}{3Q_\nu}$ =
0.024  and  0.04 g\,$cm^{-2}$  at  60  and  100 \micron\  respectively
\citep{hildebrand1983}.   The  estimated dust  masses  at  60 and  100
\micron\,  are  9.5$\times$10$^{-5}$  and  9.2$\times$10$^{-5}$~\msun\
respectively    at    an    adopted    distance    of    450~pc    and
2.07$\times$10$^{-5}$,  and  2.01$\times$10$^{-5}$~\msun\  at  210~pc.
Assuming  a  gas-to-dust  ratio   of  100,  the  estimated  disk  mass
(dust+gas)  would  be in  the  range  of  0.002 to  0.009~\msun.   The
estimated values of  the disk mass of \snm\ are  lower compared to the
disk mass  of actively accreting Herbig  Ae type stars  such as AB~Aur
(0.01  \msun\,),  HD~163296 (0.028  \msun\,)  derived from  millimetre
observations  \citep{mannings1997}.   This   is  consistent  with  the
relatively evolved circumstellar environment of \snm.
\subsection{Accretion and wind models}
The \ha\ emission line serves  as one of the most empirical signatures
of accretion  of the circumstellar material onto  the central \ac{PMS}
stars.   Magnetospheric  accretion  models  are  quite  successful  in
explaining  the  observed   spectral  characteristics  such  as  broad
asymmetric  line profiles etc.  of accreting  classical T  Tauri stars
(CTTS)   i.e.     the   lower   mass    counterparts   of   \ac{HAeBe}
\citep{hartman1994,muzerolle98model,muzerolle01p2}.     According   to
magnetospheric accretion  models, the stellar  magnetic field disrupts
the  circumstellar disk  at several  stellar radii  and  the accreting
material falls onto the stellar surface along the magnetic field lines
\cite{koenigl91}.   The  results  of magnetospheric  accretion  models
suggest that the  hydrogen emission lines form in  the infall zone, so
that the  blue-shifted asymmetric emission-line profiles  arise due to
partial obscuration  of the  flow by the  inner part of  the accretion
disk  and the  red-shifted absorption  profiles result  from infalling
material  at  near  free-fall  velocities  onto  the  stellar  surface
\citep{muzerolle98model}. \cite{muzerolle2004b}  also demonstrated the
magnetically channelled  disk accretion  in the Herbig~Ae  star UX~Ori
and partly reproduced the double-peak \ha\ profile of UX~Ori.  Several
studies e.g. \cite{alencar2005,kurosawa2006} found that disk wind also
play an  important role  in the formation  of double-peak  profiles in
CTTS.

\snm\  showed   varieties  of   \ha\  line  profiles   throughout  our
observation. The shape of  the profiles contains information about the
geometrical  orientation  and   physical  processes  involved  in  the
interaction  of  the star  with  its  circumstellar environment.   The
common  trends in  all spectra  are the  slightly  red-shifted central
absorption  and the  elongated blue-shifted  emission  components. The
width  of the  central absorption  changed within  a day.   Though the
strength of the blue-shifted  emission component varied with time, but
there  was no  significant variation  in the  maximum velocity  at the
wings.  On  the contrary,  the  red-shifted  component showed  drastic
variations  in shape,  strength,  and velocity.  The  strength of  the
red-shifted  emission,  if present,  was  less  than the  blue-shifted
emission, which can  be attributed to the selective  absorption of the
red-shifted flux  due to the asymmetric distribution  of the star-disk
system with respect to our line of sight.

\cite{kurosawa2006} investigated  the formation of  \ha\ emission line
profiles  around  CTTS  combining  the  magnetospheric  accretion  and
disk-wind  models  (hybrid  model).   Though the  \ha\  emission  line
profiles are  calculated with typical  parameters of CTTS, but  it has
also  been suggested that  the model  can be  used for  other \ac{PMS}
stars such as  \ac{HAeBe} stars and for Brown  dwarfs (BDs).  In order
to understand the relative importance of the accretion and wind on the
formation  of the line  profile \citeauthor  {kurosawa2006} calculated
the  \ha\  line profiles  for  three  cases  e.g.  the  magnetospheric
accretion,  the  disk  wind,   and  a  combination  of  magnetospheric
accretion and disk wind models. With the use of this hybrid model they
were able to reproduce the  variety of the \ha\ emission line profiles
reported   by  \cite{reipurth1996}.    \citeauthor{kurosawa2006}  also
noticed  the  degeneracy  in  the  line  profiles  that  emerged  from
different models mostly from the disk wind model and the hybrid model.
For  example, the  double  peak  \ha\ emission  line  profiles can  be
reproduced using  the disk-wind model only.  It  is somewhat difficult
to distinguish  the contribution of magnetospheric  accretion and disk
winds in the emission line profiles .  \cite{alencar2005} reported the
line  profile  variability  of   CTTS  RW~Aur  and  found  significant
variations  in the  intensity of  the blue-  and  red-shifted emission
components  similar  to  \snm  .   \citeauthor{alencar2005}  tried  to
reproduce  the  double-peaked \ha\  emission  line  profiles with  the
magnetospheric accretion model and the disk wind model.  The \ha, \ha,
and \ndo\  profiles calculated using  the disk wind models  provided a
better  fit  to  the  observed  spectra.   These  studies  reveal  the
importance of disk wind in  the emission line profiles of the \ac{PMS}
stars.  While the  magnetospheric accretion model produces double-peak
profile  under certain conditions  such as  accretion rates  etc., the
wind  model produces  double-peak profiles  in usual  conditions (e.g.
\cite{alencar2005,kurosawa2006}).   So it  is possible  that  the disk
wind  contributes significantly in  the \ha\  emission of  \snm, which
showed a double-peak profile most  of the time during our observation.
However, we  did not detect [O~I]~$\lambda$6300~\ang\  emission in our
observed spectra, which according to several authors forms in the disk
wind \citep{hartigan1995}.  However,  \cite{acke2005} proposed that [O
~I]~$\lambda$6300~\ang\ emission  arises in the surface  layers of the
protoplanetary  disks  surrounding \ac{HAeBe}  stars  due  to the  the
photodissociation of  OH molecules.   The maximum equivalent  width of
the \ha\ emission in our observed spectra is $<$ 1~\ang, which is less
than the  typical minimum  equivalent width of  the \ha\  emission, of
$\sim$4~\ang\ of  the observed samples of \ac{HAeBe}  stars studied by
\cite{acke2005}.  Considering  the discrepancy regrading  a thermal or
non-thermal origin  of the [O~I]~$\lambda$6300~\ang\  emission and the
low \ha\ emission, the absence of the line may not contradict with the
presence of a disk wind.   However, considering the rapid variation in
the \ha\ emission on timescale of a day, which can be attribute to the
variation of  wind acceleration rate  \citep{kurosawa2006}, we suggest
that the disk wind may be responsible for most of the \ha\ emission of
\snm. However, the degeneracy of accretion and disk-wind models cannot
be explored completely with our observed spectra.

We try  to interpret the behaviour  of \snm\ in the  context of hybrid
model, i.e.  by considering the effect of  magnetospheric accretion as
well as  the disk-wind.   The central absorption  that was  present in
most  of  the spectra  seemed  to  arise  because of  the  geometrical
configuration of  the star-disk system  which is viewed  viewed nearly
edge-on.   The  changes  in  the  shapes  of  the  central  absorption
component,  for  most  of  the   epochs,  seemed  to  arise  from  the
superposition of infall  or outflow events in the  same line of sight.
On  27~October~2008, the  \ha\  emission line  profile  may have  been
powered by  a disk wind with  the simultaneous presence  of infall and
outflow.  The   infall  was  quite  prominent  and   it  obscured  the
red-shifted  emission component.  Before  30~October~2008, the  infall
event ended  and the central  absorption returned to the  normal shape
with   a   nearly   symmetric   double-peak   profile.   On   28   and
29~December~2008, the  \ha\ emission  line profiles were  dominated by
the  infall event.  During  17 to  20~January~2009, the  \ha\ emission
line profiles  were dominated  by the disk  wind but  the blue-shifted
emission was  partially obscured by the prominent  outflow event.  The
\ha\ emission line profiles  observed during March and April~2009 also
dominated by  the disk  wind.  The \ha\  emission from \snm\  seems to
originate in  disk wind  and at  the same time  in sporadic  infall of
matter  moderated by a  magnetic field,  and the  outflow of  cool gas
occurs quite frequently.
\section{Conclusion}
\label{section:con}
Spectral analysis of 45  high-resolution spectra reveals that the line
profile variability of \snm\ is ubiquitous. The line profile variation
originate due to the  disk wind, rate  of wind acceleration,  episodic accretion,
and cool  outflowing gaseous  material.  These signatures  support the
complex and  dynamical interaction of the  circumstellar material with
\snm.   The  circumstellar environment  of  the  Herbig~Ae star  \snm\
consists of  an inner cavity with  gaseous material, a  dusty disk and
disk wind.   The interaction processes are  highly time-dependent from
one hour for the changes in \ac{TACs} to one day and longer timescales
of the order of months  for the changes in the overall characteristics
of the  broad \ha\ profiles. Owing  to the extensive  time coverage of
the observation down  to timescales of hours we  could follow the time
evolution of  the individual  \ac{TACs} unambiguously.  The  long term
monitoring of the star at timescales of months confirmed the deviation
of  the  \ha\  emission  line  profiles from  inverse  P~Cygni  types.
Accretion and  outflow events  are not continuous  in nature,  and the
profiles of the TACs seem to  be created by obscuration due to gaseous
blobs, which  indicate that the circumstellar environment  of \snm\ is
inhomogeneous  and  clumpy  in  nature.   Detectable  changes  in  the
\ac{TACs} showed  a deceleration of the  order of a fraction  to a few
\ms. Variation  in the blue- and red-shifted  emission components also
occurred frequently on the shortest timescale of hours, which suggests
the rate  of the wind  acceleration is also  time-dependent.  Finally,
the  presence of  emission in  the absence  of veiling  emphasises the
importance of  the disk wind  in \snm.  Models containing  dynamic and
non-axysymmetric magnetospheric  accretion with  the disk wind  may be
able   to  provide   a   satisfactory  explanation   of  the   complex
circumstellar activities of \ac{PMS} stars.

\begin{acknowledgements}
 We would like  to thank the anonymous referee  for her/his insightful
comments which helped to improve  the presentation of the paper.  This
research  has made  use  of the  SIMBAD  data base,  operated at  CDS,
Strasbourg, France.
\end{acknowledgements}
\bibliographystyle{aa}

\begin{thebibliography}{43}
\expandafter\ifx\csname natexlab\endcsname\relax\def\natexlab#1{#1}\fi

\bibitem[{{Acke} {et~al.}(2005){Acke}, {van den Ancker}, \&
  {Dullemond}}]{acke2005}
{Acke}, B., {van den Ancker}, M.~E., \& {Dullemond}, C.~P. 2005, \aap, 436, 209

\bibitem[{{Alencar} {et~al.}(2005){Alencar}, {Basri}, {Hartmann}, \&
  {Calvet}}]{alencar2005}
{Alencar}, S.~H.~P., {Basri}, G., {Hartmann}, L., \& {Calvet}, N. 2005, \aap,
  440, 595

\bibitem[{{Balona} {et~al.}(2002){Balona}, {Koen}, \& {van Wyk}}]{balona2002}
{Balona}, L.~A., {Koen}, C., \& {van Wyk}, F. 2002, \mnras, 333, 923

\bibitem[{{de Winter} {et~al.}(1999){de Winter}, {Grady}, {van den Ancker},
  {P{\'e}rez}, \& {Eiroa}}]{dewinter1999}
{de Winter}, D., {Grady}, C.~A., {van den Ancker}, M.~E., {P{\'e}rez}, M.~R.,
  \& {Eiroa}, C. 1999, \aap, 343, 137

\bibitem[{{Dunkin} {et~al.}(1997){Dunkin}, {Barlow}, \& {Ryan}}]{dunkin1997}
{Dunkin}, S.~K., {Barlow}, M.~J., \& {Ryan}, S.~G. 1997, \mnras, 286, 604

\bibitem[{{Edwards} {et~al.}(2003){Edwards}, {Fischer}, {Kwan}, {Hillenbrand},
  \& {Dupree}}]{edwards2003}
{Edwards}, S., {Fischer}, W., {Kwan}, J., {Hillenbrand}, L., \& {Dupree}, A.~K.
  2003, \apjl, 599, L41

\bibitem[{{Eiroa} {et~al.}(2001){Eiroa}, {Garz{\'o}n}, {Alberdi}, {de Winter},
  {Ferlet}, {Grady}, {Cameron}, {Davies}, {Deeg}, {Harris}, {Horne},
  {Mer{\'{\i}}n}, {Miranda}, {Montesinos}, {Mora}, {Oudmaijer}, {Palacios},
  {Penny}, {Quirrenbach}, {Rauer}, {Schneider}, {Solano}, {Tsapras}, \&
  {Wesselius}}]{eiroa2001}
{Eiroa}, C., {Garz{\'o}n}, F., {Alberdi}, A., {et~al.} 2001, \aap, 365, 110

\bibitem[{{G.~Helou \& D.~W.~Walker}(1988)}]{helou1988}
{G.~Helou \& D.~W.~Walker}, ed. 1988, {Infrared astronomical satellite (IRAS)
  catalogs and atlases. Volume 7: The small scale structure catalog}, Vol.~7

\bibitem[{{Grady} {et~al.}(1996){Grady}, {Perez}, {Talavera}, {Bjorkman}, {de
  Winter}, {The}, {Molster}, {van den Ancker}, {Sitko}, {Morrison}, {Beaver},
  {McCollum}, \& {Castelaz}}]{grady1996}
{Grady}, C.~A., {Perez}, M.~R., {Talavera}, A., {et~al.} 1996, \aaps, 120, 157

\bibitem[{{Grinin} {et~al.}(2001){Grinin}, {Kozlova}, {Natta}, {Ilyin},
  {Tuominen}, {Rostopchina}, \& {Shakhovskoy}}]{grinin2001}
{Grinin}, V.~P., {Kozlova}, O.~V., {Natta}, A., {et~al.} 2001, \aap, 379, 482

\bibitem[{{Grinin} {et~al.}(1994){Grinin}, {The}, {de Winter}, {Giampapa},
  {Rostopchina}, {Tambovtseva}, \& {van den Ancker}}]{grinin1994}
{Grinin}, V.~P., {The}, P.~S., {de Winter}, D., {et~al.} 1994, \aap, 292, 165

\bibitem[{{Guimar{\~a}es} {et~al.}(2006){Guimar{\~a}es}, {Alencar}, {Corradi},
  \& {Vieira}}]{guimaraes2006}
{Guimar{\~a}es}, M.~M., {Alencar}, S.~H.~P., {Corradi}, W.~J.~B., \& {Vieira},
  S.~L.~A. 2006, \aap, 457, 581

\bibitem[{{Hamann}(1994)}]{hamann1994}
{Hamann}, F. 1994, \apjs, 93, 485

\bibitem[{{Hamann} \& {Persson}(1992)}]{hamann1992}
{Hamann}, F. \& {Persson}, S.~E. 1992, \apjs, 82, 285

\bibitem[{{Hartigan} {et~al.}(1995){Hartigan}, {Edwards}, \&
  {Ghandour}}]{hartigan1995}
{Hartigan}, P., {Edwards}, S., \& {Ghandour}, L. 1995, \apj, 452, 736

\bibitem[{{Hartmann} {et~al.}(1994){Hartmann}, {Hewett}, \&
  {Calvet}}]{hartman1994}
{Hartmann}, L., {Hewett}, R., \& {Calvet}, N. 1994, \apj, 426, 669

\bibitem[{{Hern{\'a}ndez} {et~al.}(2005){Hern{\'a}ndez}, {Calvet}, {Hartmann},
  {Brice{\~n}o}, {Sicilia-Aguilar}, \& {Berlind}}]{hernandez2005}
{Hern{\'a}ndez}, J., {Calvet}, N., {Hartmann}, L., {et~al.} 2005, \aj, 129, 856

\bibitem[{{Hern{\'a}ndez} {et~al.}(2009){Hern{\'a}ndez}, {Calvet}, {Hartmann},
  {Muzerolle}, {Gutermuth}, \& {Stauffer}}]{hernandez2009}
{Hern{\'a}ndez}, J., {Calvet}, N., {Hartmann}, L., {et~al.} 2009, \apj, 707,
  705

\bibitem[{{Hildebrand}(1983)}]{hildebrand1983}
{Hildebrand}, R.~H. 1983, \qjras, 24, 267

\bibitem[{{Hubeny} \& {Lanz}(2000)}]{synspec2000}
{Hubeny}, I. \& {Lanz}, T. 2000, Synspec User's Guides, Version 43,
  http://nova.astro.umd.edu/Tlusty2002/pdf/syn43guide.pdf

\bibitem[{{Johns} \& {Basri}(1995)}]{john1995a}
{Johns}, C.~M. \& {Basri}, G. 1995, \aj, 109, 2800

\bibitem[{{Koenigl}(1991)}]{koenigl91}
{Koenigl}, A. 1991, \apjl, 370, L39

\bibitem[{{Kurosawa} {et~al.}(2006){Kurosawa}, {Harries}, \&
  {Symington}}]{kurosawa2006}
{Kurosawa}, R., {Harries}, T.~J., \& {Symington}, N.~H. 2006, \mnras, 370, 580

\bibitem[{{Maheswar} {et~al.}(2002){Maheswar}, {Manoj}, \&
  {Bhatt}}]{maheswar2002}
{Maheswar}, G., {Manoj}, P., \& {Bhatt}, H.~C. 2002, \aap, 387, 1003

\bibitem[{{Mannings} \& {Sargent}(1997)}]{mannings1997}
{Mannings}, V. \& {Sargent}, A.~I. 1997, \apj, 490, 792

\bibitem[{{Manoj} {et~al.}(2006){Manoj}, {Bhatt}, {Maheswar}, \&
  {Muneer}}]{manoj06}
{Manoj}, P., {Bhatt}, H.~C., {Maheswar}, G., \& {Muneer}, S. 2006, \apj, 653,
  657

\bibitem[{{Marconi} {et~al.}(2000){Marconi}, {Ripepi}, {Alcal{\'a}}, {Covino},
  {Palla}, \& {Terranegra}}]{marconi2000}
{Marconi}, M., {Ripepi}, V., {Alcal{\'a}}, J.~M., {et~al.} 2000, \aap, 355, L35

\bibitem[{{Monnier} \& {Millan-Gabet}(2002)}]{monnier2002}
{Monnier}, J.~D. \& {Millan-Gabet}, R. 2002, \apj, 579, 694

\bibitem[{{Montesinos} {et~al.}(2009){Montesinos}, {Eiroa}, {Mora}, \&
  {Mer{\'{\i}}n}}]{montesinos2009}
{Montesinos}, B., {Eiroa}, C., {Mora}, A., \& {Mer{\'{\i}}n}, B. 2009, \aap,
  495, 901

\bibitem[{{Mora} {et~al.}(2004){Mora}, {Eiroa}, {Natta}, {Grady}, {de Winter},
  {Davies}, {Ferlet}, {Harris}, {Miranda}, {Montesinos}, {Oudmaijer},
  {Palacios}, {Quirrenbach}, {Rauer}, {Alberdi}, {Cameron}, {Deeg},
  {Garz{\'o}n}, {Horne}, {Mer{\'{\i}}n}, {Penny}, {Schneider}, {Solano},
  {Tsapras}, \& {Wesselius}}]{mora2004}
{Mora}, A., {Eiroa}, C., {Natta}, A., {et~al.} 2004, \aap, 419, 225

\bibitem[{{Mora} {et~al.}(2002){Mora}, {Natta}, {Eiroa}, {Grady}, {de Winter},
  {Davies}, {Ferlet}, {Harris}, {Montesinos}, {Oudmaijer}, {Palacios},
  {Quirrenbach}, {Rauer}, {Alberdi}, {Cameron}, {Deeg}, {Garz{\'o}n}, {Horne},
  {Mer{\'{\i}}n}, {Penny}, {Schneider}, {Solano}, {Tsapras}, \&
  {Wesselius}}]{mora2002}
{Mora}, A., {Natta}, A., {Eiroa}, C., {et~al.} 2002, \aap, 393, 259

\bibitem[{{Muzerolle} {et~al.}(1998){Muzerolle}, {Calvet}, \&
  {Hartmann}}]{muzerolle98model}
{Muzerolle}, J., {Calvet}, N., \& {Hartmann}, L. 1998, \apj, 492, 743

\bibitem[{{Muzerolle} {et~al.}(2001){Muzerolle}, {Calvet}, \&
  {Hartmann}}]{muzerolle01p2}
{Muzerolle}, J., {Calvet}, N., \& {Hartmann}, L. 2001, \apj, 550, 944

\bibitem[{{Muzerolle} {et~al.}(2004){Muzerolle}, {D'Alessio}, {Calvet}, \&
  {Hartmann}}]{muzerolle2004b}
{Muzerolle}, J., {D'Alessio}, P., {Calvet}, N., \& {Hartmann}, L. 2004, \apj,
  617, 406

\bibitem[{{Natta} {et~al.}(2000){Natta}, {Grinin}, \&
  {Tambovtseva}}]{natta2000}
{Natta}, A., {Grinin}, V.~P., \& {Tambovtseva}, L.~V. 2000, \apj, 542, 421

\bibitem[{{Oudmaijer} {et~al.}(2001){Oudmaijer}, {Palacios}, {Eiroa}, {Davies},
  {de Winter}, {Ferlet}, {Garz{\'o}n}, {Grady}, {Cameron}, {Deeg}, {Harris},
  {Horne}, {Mer{\'{\i}}n}, {Miranda}, {Montesinos}, {Mora}, {Penny},
  {Quirrenbach}, {Rauer}, {Schneider}, {Solano}, {Tsapras}, \&
  {Wesselius}}]{oudmaijer2001}
{Oudmaijer}, R.~D., {Palacios}, J., {Eiroa}, C., {et~al.} 2001, \aap, 379, 564

\bibitem[{{Rao} {et~al.}(2005){Rao}, {Sriram}, {Jayakumar}, \&
  {Gabriel}}]{nkrao2005}
{Rao}, N.~K., {Sriram}, S., {Jayakumar}, K., \& {Gabriel}, F. 2005, JApA, 26,
  331

\bibitem[{{Reipurth} {et~al.}(1996){Reipurth}, {Pedrosa}, \&
  {Lago}}]{reipurth1996}
{Reipurth}, B., {Pedrosa}, A., \& {Lago}, M.~T.~V.~T. 1996, \aaps, 120, 229

\bibitem[{{Ripepi} {et~al.}(2003){Ripepi}, {Marconi}, {Bernabei}, {Palla},
  {Pinheiro}, {Folha}, {Oswalt}, {Terranegra}, {Arellano Ferro}, {Jiang},
  {Alcal{\'a}}, {Marinoni}, {Monteiro}, {Rudkin}, \& {Johnston}}]{ripepi2003}
{Ripepi}, V., {Marconi}, M., {Bernabei}, S., {et~al.} 2003, \aap, 408, 1047

\bibitem[{{van den Ancker} {et~al.}(1996){van den Ancker}, {The}, \& {de
  Winter}}]{vanancker1996}
{van den Ancker}, M.~E., {The}, P.~S., \& {de Winter}, D. 1996, \aap, 309, 809

\bibitem[{{van Leeuwen}(2007)}]{hipnew07}
{van Leeuwen}, F. 2007, \aap, 474, 653

\bibitem[{{Vieira} {et~al.}(2003){Vieira}, {Corradi}, {Alencar}, {Mendes},
  {Torres}, {Quast}, {Guimar{\~a}es}, \& {da Silva}}]{vieira2003}
{Vieira}, S.~L.~A., {Corradi}, W.~J.~B., {Alencar}, S.~H.~P., {et~al.} 2003,
  \aj, 126, 2971

\bibitem[{{Young} {et~al.}(1989){Young}, {Xie}, {Kenney}, \&
  {Rice}}]{young1998}
{Young}, J.~S., {Xie}, S., {Kenney}, J.~D.~P., \& {Rice}, W.~L. 1989, \apjs,
  70, 699

\end{thebibliography}

\end{document}